\documentclass[aps,pra,amsmath,amssymb,twocolumn,superscriptaddress,notitlepage]{revtex4-1}
\usepackage{graphicx}
\usepackage{color}
\usepackage{braket}
\definecolor{blue}{RGB}{25,51,170}
\definecolor{green}{RGB}{51, 170, 42}
\definecolor{red}{RGB}{238,25,25}
\usepackage[colorlinks,citecolor=blue,linkcolor=red,urlcolor=green]{hyperref}
\hypersetup{
  pdfkeywords={},
  pdftitle={Quantum Scanning Microscope for Cold Atoms},
  pdfauthor={D. Yang, D. V. Vasilyev, C. Laflamme, M. A. Baranov, and P. Zoller}}

\makeatletter

\begin{document}

\title{Quantum Scanning Microscope for Cold Atoms}

\author{D. Yang}

\author{D. V. Vasilyev}

\author{C. Laflamme}

\author{M. A. Baranov}

\author{P. Zoller}

\address{Center for Quantum Physics, Faculty of Mathematics, Computer Science and Physics,
University of Innsbruck, A-6020 Innsbruck, Austria}

\address{Institute for Quantum Optics and Quantum Information of the Austrian
Academy of Sciences, A-6020 Innsbruck, Austria}

\date{\today}
\begin{abstract}
We present a detailed theoretical description of a recently proposed atomic scanning
microscope in a cavity QED setup [D. Yang \emph{et al.}, Phys.~Rev.~Lett.~\textbf{120},~133601~(2018)].
The microscope continuously observes atomic densities with optical
subwavelength resolution in a nondestructive way. The super-resolution is achieved by engineering an internal atomic dark state with a sharp spatial variation of population of a ground level dispersively coupled to the cavity field. Thus, the atomic position encoded in the internal state is revealed as a phase shift of the light reflected from the cavity in a homodyne experiment.
Our theoretical description of the microscope
operation is based on the stochastic master equation describing the
conditional time evolution of the atomic system under continuous observation
as a competition between dynamics induced by the Hamiltonian of the
system, decoherence effects due to atomic spontaneous decay, and
the measurement backaction. Within our approach we relate the observed
homodyne current with a local atomic density, and discuss the emergence
of a quantum nondemolition measurement regime allowing continuous
observation of spatial densities of quantum motional eigenstates
without measurement backaction in a single experimental~run. 
\end{abstract}
\maketitle

\section{Introduction}

\label{sec1} In recent work we have proposed
and discussed a scanning quantum microscope for cold atoms to continuously
monitor atomic quantum dynamics~\cite{Yang2017a}. The unique feature of our setup is that it acts a \emph{continuous measurement quantum device} which, depending
on the mode of operation, implements an emergent quantum nondemolition
(QND) measurement of local density of an atomic quantum state with subwavelength
resolution. It is, therefore, conceptually different from the familiar
quantum gas microscope~\cite{*[{For a review, see }] [{ and references therein.}] Kuhr2016}
that takes a fluorescence image of an instantaneous distribution of
atoms over lattice sites in a many-body system placed in an optical
lattice. In present experiments, the quantum gas microscope operates
as a destructive measurement device, making continuous
observation of the atomic dynamics impossible. In contrast, our proposed microscope
does not take pixelized images of atomic lattices at a given time,
but scans in time the atomic quantum state on the subwavelength scale.
In the \emph{Movie Mode}, for a fixed focal region the microscope
continuously records the atomic wave packet dynamics. In the \emph{Scanning
Mode} with a good cavity, the microscope appears as a quantum nondemolition
device such that a \emph{single} spatial scan of the microscope
focal region maps out the spatial density of an atomic motional eigenstate.

The quantum scanning microscope~\cite{Yang2017a} continuously measures the atomic
density within its focal region of subwavelength size via dispersive
coupling of atoms to a laser driven cavity, while 
%its output field  
the light reflected from the cavity is monitored by a homodyne detection within
%and performing homodyne detection of the light reflected from the cavity, 
the framework
of weak continuous measurements~\cite{Braginsky1992,wiseman2009quantum,gardiner2015quantum}.
It builds on the idea of using the atom-cavity coupling for measurement
and control of atomic quantum systems, which was employed in experiments~\cite{PhysRevLett.109.133603,Schreppler1486}
as well as in theoretical proposals~\cite{Steck2004,Lee2014,Wade2015,Mazzucchi2016,Ashida2017,Laflamme2017}.
The microscope achieves the spatial super-resolution by entangling the internal
state of an atom with its position via engineering a spatially dependent
dark state \cite{0953-4075-39-17-002,Gorshkov2008} (see Refs.~\cite{PhysRevX.3.031014,PhysRevA.92.033838} for pioneering experiments), however optimized such that the perturbation of the atomic system by the probe is negligible.
This is in contrast to the existing methods for achieving subwavelength
resolution by correlating the position of an atom with its internal
state via either spatial potential gradients~\cite{Thomas1991,Weitenberg2011b}
or nonlinear optical response~\cite{0953-4075-39-17-002,Gorshkov2008,Maurer2010}, which typically suffer from the presence of strong forces acting on
atoms. We also note that according to Ref.~\cite{Ashida2015} advanced data processing 
makes it possible to reach a resolution comparable to the size of
vibrational ground state of atoms in optical lattice wells, but still
does not allow to ``look into'' the lattice site and to monitor
dynamics continuously.

It is the purpose of the present paper to elaborate on the detailed theory
behind the operation of the quantum scanning microscope for cold atoms beyond the short presentation
in Ref.~\cite{Yang2017a}, with emphasis on decoherence effects
caused by atomic spontaneous emission, and addressing experimental
feasibility of the scheme. This also includes a thorough analysis
of the effects of measurement backaction, the emergent quantum
nondemolition regime, and the microscope resolution limit. The paper is organized as follows. In Sec.~\ref{sec2},
we discuss the cavity QED (CQED) setup and operation principles of
the microscope. The stochastic master equation (SME) describing the
microscope operation will be derived in Sec.~\ref{sec:sec2} starting
from a quantum optical model for a CQED setup including the atomic
spontaneous emission. Based on this derivation, in Sec.~\ref{sec4}
we present a detailed analysis of the \textit{Movie} and the\textit{ Scanning} operation
modes of the microscope illustrated by several (physically interesting)
examples. We discuss the experimental feasibility of the proposed 
setup in Sec.~\ref{sec:exp_considerations} and conclude in Sec.~\ref{sec:conclusion}.

\section{Microscope setup and operation principles}
\label{sec:operationPrinciples}
\begin{figure}[t!]
\centering{}\includegraphics[width=0.49\textwidth]{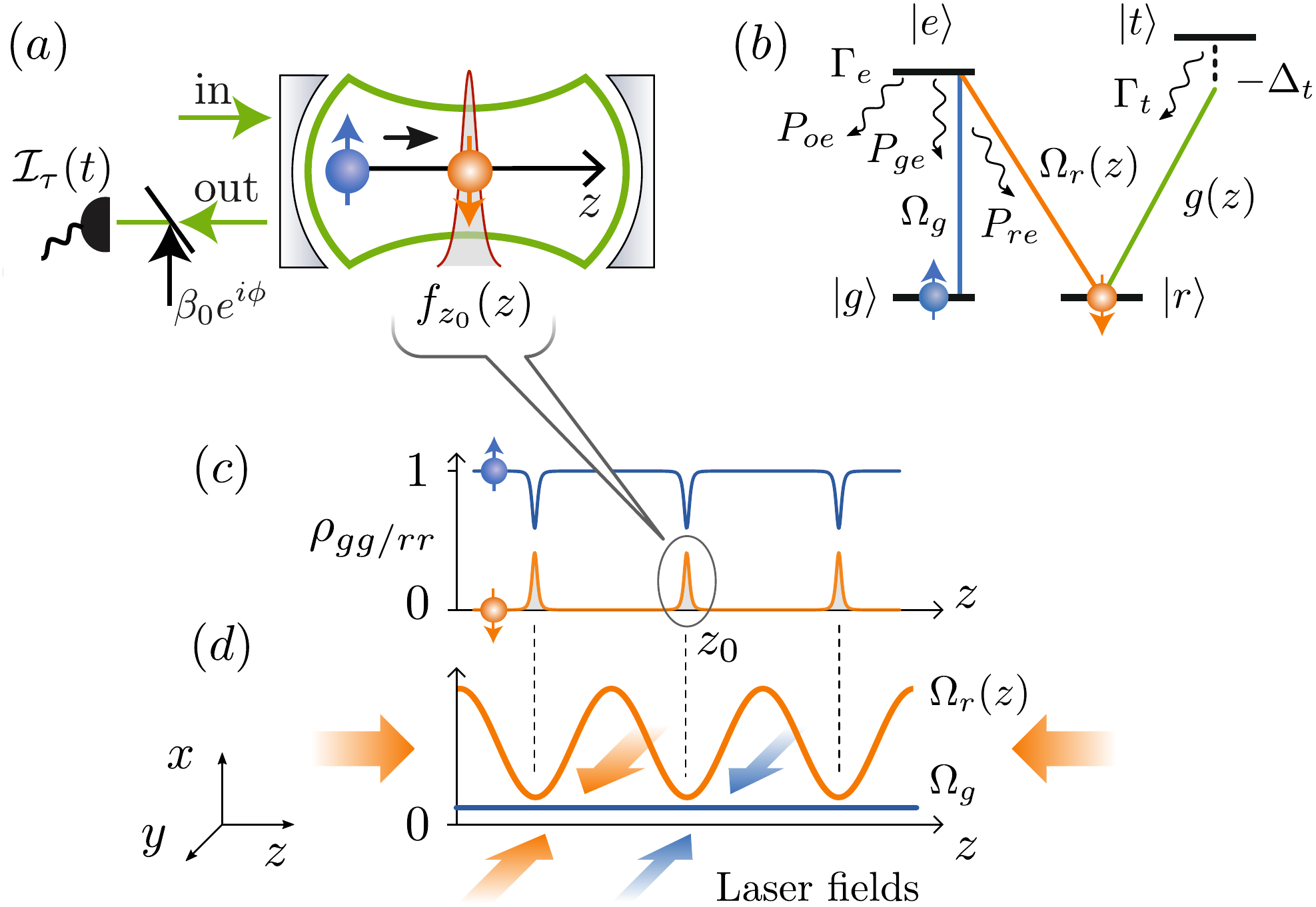} \caption{Overview of the proposed Quantum Scanning Microscope. (a) The appearance of an atom in the focal region of the microscope, as defined by the \emph{focus
function} $f_{z_{0}}(z)$, is accompanied by an internal
spin flip, which shifts the resonance of a cavity mode and can be
detected via homodyne detection. (b) Atomic level structure used to
realize the microscope. The $\Lambda$ system $\ket{g},\ket{e},\ket{r}$ supports a dark state $\ket{D(z)}$, of which the spin structure is correlated with the position of the atom due to the spatially varying Rabi frequency $\Omega_r(z)$. This internal spin is detected
via dispersively coupling the $\ket{r}\to\ket{t}$ transition to a
cavity mode. Atomic spontaneous emission, as shown by the wavy lines, results in imperfections in the microscope performance. Here $\Gamma_{e(t)}$ are the spontaneous emission rate for the level $\ket{e}(\ket{t})$, and $P_{je}$ (with $j={g,r,o}$) denotes the branching ratio for the emission channel $\ket{e}\to\ket{j}$, with $\ket{o}$ denoting atomic levels outside the four-level system under consideration. (c) The subwavelength spin structure of the atomic dark state $\ket{D(z)}$, and (d) the corresponding laser configuration (see description in Sec.~\ref{sec2} and~\ref{sec:focusingFunction}).}
\label{fig:figure1} 
\end{figure}

\label{sec2} We find it worthwhile to start our discussion with a summary of the microscope operation principles, before entering technical details. In the present paper we will focus on single particle experiments illustrating the main concepts, which, however, are immediately applicable to many-body systems~\cite{Yang2017a}. The basic idea behind the quantum scanning microscope is to
entangle, with subwavelength resolution, the position of an atom with
its internal state, such that the observation of the internal state
provides information about the atomic position. In the proposed setup (cf.~Fig.~\ref{fig:figure1} and~\cite{Yang2017a}), the entanglement
is achieved by using a position-dependent dark state in an atomic
$\Lambda$ system, which allows to achieve an
internal state flip in a region of an optical subwavelength size (the focal
region of the microscope)\cite{Gorshkov2008,Lacki2006,Gorshkov2016,0953-4075-39-17-002,PhysRevX.3.031014,PhysRevA.92.033838}. Via a dispersive coupling
of one of the internal atomic state to a cavity mode, the change in
the internal state of an atom entering the focal region is detected
nondestructively as a shift in the mode resonance frequency, which
is revealed as a phase shift of the laser light reflected from the
cavity in a homodyne measurement.

To be more specific, we consider an atom (or system of atoms) with
two internal ground (spin) states $\ket{g}$ and $\ket{r}$, and one
excited state $\ket{e}$ moving along the $z$-axis, see Fig.~\ref{fig:figure1},
and we assume strong confinement in the other directions. The Hamiltonian describing
a 1D atomic motion is 
\begin{equation}
\hat{H}_{{A,E}}=\frac{\hat{p}_{z}^{2}}{2m}+V(\hat{z}),\label{eq:Hext}
\end{equation}
where $V(\hat{z})$ is an external (off-resonant) optical potential
constraining the motion alone the $z$ axis, which we assume to be
identical for all atomic internal states. To entangle the position
of an atom with its internal state, we form a $\Lambda$ system
with two Rabi frequencies: a constant weak $\Omega_{g}$ and a strong
position dependent (periodic) $\Omega_{r}(z)$ indicated in Figs.~\ref{fig:figure1}(b)
and (d) by the blue and the orange color, respectively. Note that,
in contrast to Refs.~\cite{Lacki2006,Gorshkov2016}, here the Rabi Frequency
$\Omega_{r}(z)$ is never zero, $\Omega_{r}(z)>0$. This configuration,
as explained in detail in Sec.~\ref{sec:focusingFunction}, makes it possible
to create a dark state 
\begin{equation}
\label{eq: dark_state_def0}
\ket{D(z)}\sim\Omega_r(z)\ket{g} -\Omega_g\ket{r},
\end{equation}
with a subwavelength spatial
structure \textit{without} generating a noticeable nonadiabatic potential barrier, thus minimizing backaction.
In the dark state, see Figs.~\ref{fig:figure1}(c), the internal
state $\ket{r}$ is partially populated only in the narrow focal regions
of size $\sigma\ll\lambda_{0}$ near the minima of $\Omega_{r}(z)$, which creates
the desired internal-state--position entanglement. 

To detect an atom in the internal state $\ket{r}$ and, therefore,
inside the focal region, we place the atomic system into a laser driven
optical cavity, see Fig.~\ref{fig:figure1}(a), such that the driven
cavity mode {[}the green area in Fig.~\ref{fig:figure1}(a) and the
green line in Fig.~\ref{fig:figure1}(b){]} couples the state $\ket{r}$
to another excited atomic state $\ket{t}$ with detuning $\Delta_{t}$
and strength $g(z)$ (the $z$-dependence is due to a spatial profile
of the cavity mode). For a large detuning, $\left|\Delta_{t}\right|\gg\left|g(z)\right|$,
this coupling generates a local dispersive interaction between the atom and the cavity mode. As detailed in Sec.~\ref{sec:elimination_internal}, this interaction can be written as 
\begin{equation}
\hat{H}_{{\rm coup}}=\mathcal{A}f_{z_{0}}(\hat{z})\,\hat{c}^{\dagger}\hat{c},\label{eq:dispersive_coup}
\end{equation}
where $\hat{c}^{\dagger}$($\hat{c}$) is the photon creation (annihilation)
operator for the cavity mode, and

\begin{equation}
\mathcal{A}f_{z_{0}}(\hat{z})=\frac{\hbar g^{2}(\hat{z})}{\Delta_{t}}\left|\left\langle r|D(z)\right\rangle \right|^{2}=\frac{\hbar g^{2}(\hat{z})}{\Delta_{t}}\rho_{rr}\label{eq:dispersive_coup-1}
\end{equation}
defines a sharply peaked \emph{focusing function} $f_{z_{0}}(\hat{z})$ of \emph{resolution}
(width) $\sigma$ around the \emph{focal point} $z_{0}$ {[}the minimum
of $\Omega_{r}(z)${]}, see Figs.~\ref{fig:figure1}(c-d).
Here $\mathcal{A}$ is the coupling strength with the dimension of
energy, chosen in such a way that the matrix element of the focusing
function over the atomic motional eigenstates are of order $1$ (the
precise definition will be given below in Sec.~\ref{sec:sec2}, together
with discussion of the effects related to a finite life time of the
levels $\ket{t}$ and $\ket{e}$).

The dispersive coupling \eqref{eq:dispersive_coup} implies that the
presence of an atom inside the \emph{focal region} defined by $f_{z_{0}}(z)$
shifts the cavity resonance, which can be detected via homodyne measurement.
In such a measurement the output field of the cavity is combined with
a local oscillator with phase $\phi$, resulting in a homodyne current
of the form (see Sec.~\ref{sec:sec2})
\begin{equation}
I(t)=\sqrt{\kappa}\langle\hat{X}_{\phi}\rangle_{c}+\xi(t).\label{eq:homodyne_review}
\end{equation}
Here, $\hat{X}_{\phi}\equiv e^{i\phi}\hat{c}^{\dagger}+e^{-i\phi}\hat{c}$
is the associated quadrature operator of the cavity mode, $\kappa$
is the cavity damping rate, $\xi(t)$ is the shot noise of the electromagnetic
field, and $\langle\dots\rangle_{c}\equiv\mathrm{Tr}\{\ldots\,\rho_{c}(t)\}$
refers to an expectation value with respect to the density matrix
of the joint atom-cavity system, conditioned on the measurement outcome.
The evolution of $\rho_{c}(t)$ is governed by a stochastic master
equation (SME) to be derived in Sec.~\ref{sec:sec2}. Based on this equation
we present a detailed discussion of the two operation modes of the
microscope, the \textit{Movie Mode} (Sec.~\ref{badCavityLimit}) and the \textit{Scanning
Mode} (Sec.~\ref{sec3}), and establish in both cases the connection
between the measured homodyne current and the atomic motional quantum
state.

In the \textit{Movie Mode}, the microscope is focused at a given (fixed) position
$z_{0}$ to ``record a movie'' of an atomic wave packet passing through the observation region, which in the example discussed below will be illustrated by a coherent state in a harmonic potential $V(z)$.
This requires the cavity response
time $\tau_{C}=1/\kappa$ being much smaller than the typical timescale
$1/\omega$ associated with the atomic motion, where $\omega$ is an oscillation frequency, i.e.~we are in the \emph{bad cavity limit} $\kappa\gg\omega$. In this case,
as shown in Sec.~\ref{badCavityLimit}, the homodyne current follows
the expectation value of the focusing function $\langle f_{z_{0}}(\hat{z})\rangle_{c}=\mathrm{Tr}\{f_{z_{0}}(\hat{z})\tilde{\rho}_{c}(t)\}$ instantaneously
with   $\tilde{\rho}_{c}(t)$ the \emph{atomic} conditional density matrix,
\begin{equation}
I(t)=2\sqrt{\gamma}\langle f_{z_{0}}(\hat{z})\rangle_{c}+\xi(t) .\label{eq:Homodyne_movie}
\end{equation}
Here $\gamma=[4\mathcal{A}\mathcal{E}/(\hbar\kappa)]^{2}$ is the
effective coupling rate for the measurement with $\mathcal{E}$ 
the amplitude of the driving laser field. Therefore, the measured
homodyne current $I(t)$ in this mode directly probes the time evolution
of the local atomic density at $z_{0}$ with spatial resolution $\sigma$.
We note that the measurement backaction is proportional to
$\gamma$ (detailed in Sec.~\ref{badCavityLimit}), while the signal strength
is to proportional to $\sqrt{\gamma}$, see Eq.~(\ref{eq:Homodyne_movie}). As a consequence,
we can minimize the backaction by taking small $\gamma$ and obtain
a good signal-to-noise ratio (SNR) by averaging over repeated runs
of the experiment.

In the \textit{Scanning mode} we consider the \emph{good cavity limit} $\kappa\ll\omega$
and  perform a \emph{slow scan} of the focal point $z_{0}\equiv z_{0}(t)$
across a spatial region of interest.
In this case, the cavity operates effectively as a low-pass filter for both
the measured photocurrent and the vacuum fluctuations of the electromagnetic field perturbing the atomic system under observation. As a result (see Sec.~\ref{sec3}), the homodyne current traces the atomic dynamics at $z_{0}$
averaged over many oscillation periods, such that the current is related only to the diagonal part
$\hat{f}_{z_{0}}^{(0)}=\sum_{n}|n\rangle\langle n|f_{z_{0}}(\hat{z})|n\rangle\langle n|$
of the focusing function in the basis of the eigenstates $|n\rangle$
of the motional Hamiltonian $\smash{\hat{H}_{{A,E}}}$, Eq.~(\ref{eq:Hext}). The vacuum fluctuations also couple mainly to ${\hat{f}}_{z_{0}}^{(0)}$ and, hence, do not interfere with the measurement, thus leading to a high SNR. The overall effect can be described as \emph{emergence} of a new observable $\hat{f}_{z_{0}}^{(0)}$ which commutes
with $\hat{H}_{{A,E}}$ and, therefore, represents a \textit{QND observable}
allowing for continuous quantum measurement without backaction~\cite{Gleyzes2007,Johnson2010,Volz2011,Barontini2015,Moller2017}.
Thus the microscope in the Scanning Mode appears as an effective QND
device.  Below we show (see
Sec.~\ref{sec3}) that a {\em single scan} of the microscope with spatial subwavelength resolution $\sigma$ will initially collapse (on a fast time scale $\sim 1/\gamma$) the atomic state to one of the motional eigenstates $\ket{n}$. The following  (slow) spatial scan the will map out the spatial density of $|n\rangle$. The scan of this spatial density will be reflected in the homodyne current
\begin{equation}
I(t)=2\sqrt{\gamma}\langle n|\hat{f}_{z_{0}(t)}^{(0)}|n\rangle+\xi(t).\label{eq:Homodyne_Scan}
\end{equation} 

We wish to elaborate briefly on the physics of the {\em emergent QND measurement}. We first note that the spatially
localized focusing function $f_{z_{0}}(\hat{z})$ does \textit{not} commute with the atomic motional Hamiltonian $\hat{H}_{{A,E}}$
and, therefore, is not an \textit{a priori} QND observable. In the good cavity regime,
however, the homodyne current probes only the diagonal part $\hat{f}_{z_{0}}^{(0)}$
of $f_{z_{0}}(\hat{z})$, which becomes the \emph{emergent} QND observable. In fact, this is valid for arbitrary observable, see Ref.~\cite{Yang2017a}, and was recently used in a proposal for measuring the number of atoms via a dispersive coupling to a good cavity~\cite{Uchino2018}.

While the discussion in the present paper will focus on the theory of the  quantum scanning microscope for (motion of) single particles, these concepts  generalize to many-body quantum systems.  This was illustrated in \cite{Yang2017a} (see also Ref.~[11] therein) with the example of Friedel oscillations caused by an impurity in a Fermi gas.

\section{Quantum Optical Model}
\label{sec:sec2}
In this section we describe the quantum optical model for the scanning microscope of Sec.~\ref{sec2}.  We will formulate our model in the language of a quantum stochastic Schr\"odinger equation (QSSE)  (see, e.g., Chap.~9 in Ref.~\cite{gardiner2015quantum} for an introduction), which describes the evolution of the joint atom-cavity system interacting with an external electromagnetic field environment (Sec.~\ref{sec:fullHamiltonian}).  In Sec.~\ref{sec:SME_full} we take this QSSE as the starting point to derive the SME for continuous homodyne detection of the cavity output field (see, e.g., Chap.~20 in Ref.~\cite{gardiner2015quantum} for an introduction).  By further eliminating the atomic  internal degrees of freedom (DOFs), we arrive in Sec.~\ref{sec:elimination_internal} at a SME describing the dynamics of the microscope. Furthermore, we discuss the engineering of the subwavelength focusing function and the signal filtering for homodyne detection in Sec.~\ref{sec:focusingFunction} and Sec.~\ref{sec: sigfiltering}, respectively.

\subsection{Quantum Stochastic Schr\"odinger Equation}
\label{sec:fullHamiltonian}
\begin{figure}[t]
\centering{}\includegraphics[width=0.42\textwidth]{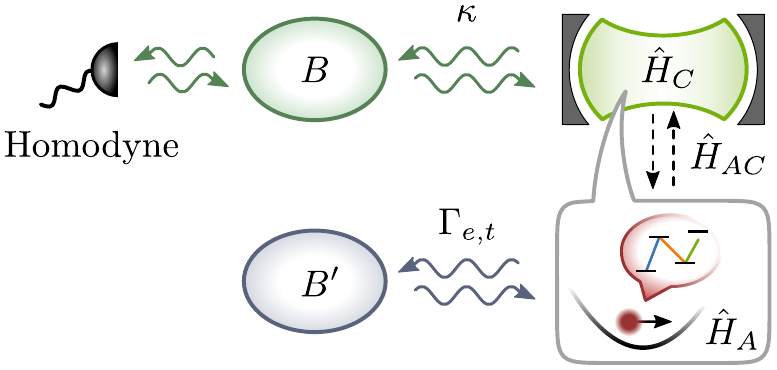} \caption{Schematics of the quantum optical model for the microscope. It consists of the atomic system which we would like to measure (described by the Hamiltonian $\hat{H}_A$), the cavity mode which enables the measurement (described by the Hamiltonian $\hat{H}_C$ and coupled to the atom via the Hamiltonian $\hat{H}_{AC}$), and external baths ($B$ and $B'$). Bath $B$ is a 1D electromagnetic field (optical fiber), which couples to the cavity mode and is under continuous homodyne detection. Bath $B'$ is a 3D electromagnetic field, which couples to the atom and gives rise to the atomic spontaneous emission.}
\label{fig:figure2} 
\end{figure}

We consider the  quantum optical model of the quantum scanning microscope as shown schematically in Fig.~\ref{fig:figure2}. The time evolution of the total system is described by the (It\^o) QSSE~\cite{wiseman2009quantum,gardiner2015quantum} for the atom-cavity system and the external electromagnetic field (bath DOFs),
\begin{align}
\label{eq:QSSEfull}
d |\Psi \rangle  &= - \left[\frac{i}{\hbar} \hat{H}_{S}+\frac{\kappa}{2} \hat{c}^\dagger \hat{c}+\frac{1}{2}(\Gamma_t\hat{\sigma}_{tt}+\Gamma_e\hat{\sigma}_{ee})\right] \ket{\Psi}dt\nonumber\\
&+\sqrt{\kappa} \hat{c}\, d\hat{B}^\dagger(t) |\Psi\rangle\notag\\
&+\!\mathop{\sum_{n=e,t}}_{j=g,r}\int \! du \sqrt{\Gamma_nP_{jn}N_{jn}(u)}\hat{\sigma}_{jn}e^{-ik_0 u\hat{z}} d\hat{B}^\dag_{jn}(u,t)\!\ket{\Psi},
\end{align}
Here, the first line includes the coherent evolution under the atom-cavity Hamiltonian $\hat{H}_S$, the cavity damping and the atomic decay, while the second and third lines represent the cavity-bath coupling and atomic spontaneous emission including recoil, respectively. Below we go through each term of Eq.~\eqref{eq:QSSEfull} in detail.

We start with $\hat{H}_{S}$, which consists of three parts,
\begin{equation}
\hat{H}_{S}=\hat{H}_A+\hat{H}_{AC}+\hat{H}_{C}.
\end{equation}
Here, $\hat{H}_A = \hat{H}_{A,E}+\hat{H}_{A,I}$ is the atomic Hamiltonian with $\hat{H}_{A,E}$ describing the external motion of the atom in 1D [cf.~Eq.~\eqref{eq:Hext}], and $\hat{H}_{A,I}$ representing the internal structure of the atom [see Fig.~\ref{fig:figure1}(b)], which in a rotating frame is given by
\begin{align}
\hat{H}_{A,I} =&-\hbar\Delta_{t}\hat{\sigma}_{tt}+\frac{\hbar }{2} \Big[ \Omega_g\hat{\sigma}_{ge} + \Omega_r(z)\hat{\sigma}_{re} + \mathrm{H.c.} \Big].
\end{align}
Here we have adopted the notation $\hat{\sigma}_{ij}\equiv\ket{i}\bra{j}$.  For simplicity, we assume a resonant drive on both the $\ket{g}\rightarrow\ket{e}$, and $\ket{r}\rightarrow\ket{e}$ transition, and thus exact Raman resonance, while the transition $\ket{r}\rightarrow\ket{t}$ is coupled to the cavity mode with off-resonant detuning
$\Delta_t=\omega_C-\omega_{tr}$, where $\omega_C$ is the frequency of the cavity mode. 

The coupling between the cavity mode and the atomic transition $\ket{r}\to\ket{t}$ is described by the Hamiltonian
\begin{equation}
\label{eq:JCcoupling}
\hat{H}_{{AC}}=\hbar g(\hat{z})\left(\hat{c}^\dag \hat{\sigma}_{rt}+\mathrm{H.c.}\right),
\end{equation}
where $\hat{c} $ $(\hat{c}^\dagger)$ is the destruction (creation) operator of the cavity mode, and $g(z)$ is the coupling strength determined by the spatial profile of the cavity mode.

The cavity mode is \emph{resonantly} driven by a classical laser beam with amplitude $\mathcal{E}$ (assumed real for simplicity), as described by the Hamiltonian $\hat{H}_C$ in the rotating frame,
\begin{equation}
\hat{H}_{C}=i\hbar\sqrt{\kappa}\mathcal{E}(\hat{c}-\hat{c}^{\dagger}),
\label{eq:cavity_intro}
\end{equation}
with $\kappa$ the damping rate of the cavity mode. 

The cavity is coupled to a waveguide (optical fiber) representing the input and output channels of our system. These external electromagnetic modes are modelled as a 1D bosonic bath (shown as bath $B$ in Fig.~\ref{fig:figure2}), and quantum optics introduces corresponding bosonic noise operators $\hat{b}(t)$ and $\hat{b}^\dagger (t)$, satisfying  white noise commutation relations $[\hat{b}(t), \hat{b}^\dag(t^\prime)]=\delta(t-t^\prime)$~\cite{wiseman2009quantum,gardiner2015quantum}. In the It\^o QSSE \eqref{eq:QSSEfull} these noise operators are transcribed as Wiener operator noise increments, $\hat{b}(t)dt\rightarrow d\hat{B}(t)$ and $\hat{b}^\dagger(t)dt\rightarrow d\hat{B}^\dagger(t)$. With the incident coherent field driving the cavity  already transformed into the classical field in Eq.~\eqref{eq:cavity_intro}, we can assume vacuum inputs, and thus have the It\^o table~\cite{wiseman2009quantum,gardiner2015quantum},
\begin{equation}
\label{eq:vacuumItoLaw1}
\begin{split}
d\hat{B}(t) d\hat{B}^\dag(t)&= dt,\\
d\hat{B}^\dag(t)d\hat{B}(t)&=d\hat{B}(t) d\hat{B}(t)= d\hat{B}^\dag(t) d\hat{B}^\dag(t)=0.
\end{split}
\end{equation}
In this formalism the cavity coupling to the waveguide  is now described by the second line of Eq.~\eqref{eq:QSSEfull} \footnote{The absence of the adjoint term is again due to our assumption $\sqrt{\kappa} \hat{c}^\dag\, d\hat{B}(t) \ket{\Psi}=0$ of vacuum inputs.}.
 Apart from such an explicit cavity-bath coupling term, the inclusion of the 1D bosonic bath also introduces a cavity damping term $-\kappa\hat{c}^\dag\hat{c}/2$ in the first line of Eq.~\eqref{eq:QSSEfull}. Mathematically, this non-Hermitian term appears as {\em It\^o -correction}, when transforming the Stratonovich QSSE to  It\^o form~\cite{wiseman2009quantum,gardiner2015quantum}. 

The third line of Eq.~\eqref{eq:QSSEfull} represents spontaneous emission of the atom into the 3D background electromagnetic modes (shown as bath $B'$ in Fig.~\ref{fig:figure2}), as familiar from the theory of laser cooling~\cite{gardiner2015quantum}. Here, $\Gamma_n$ (with $n=e,t)$ is the spontaneous emission rate of the excited states [see Fig.~\ref{fig:figure1} (b)], and $P_{jn}$ (with $j=g,r)$ denotes the branching ratio for the emission channel $\ket{n}\to\ket{j}$. The function $N_{jn}(u)$ reflects the dipole emission pattern of channel $\ket{n}\to\ket{j}$ which, for the 1D atomic motion considered here, depends on a single variable $u\equiv\cos\varphi\in[-1,1]$ with $\varphi$ the angle between the wavevector of the emitted photon and the $z$ axis. The spontaneous emission is accompanied by the momentum recoil to the 1D atomic motion, which is accounted by the operator $e^{i k_0 u\hat{z}}$ with $k_0$ the wavevector of the emitted photons (for simplicity of notation assumed to be the same for all the emission channels). For each emission channel $\ket{n}\to\ket{j}$ and for each emission direction $u$, we introduce the corresponding quantum noise increment $d\hat{B}^\dag_{jn}(u,t)$ to describes the relevant electromagnetic modes. Assuming a 3D bath initially in the vacuum state, they obey the  It\^o table~\cite{gardiner2015quantum},
\begin{equation}
\label{eq:vacuumItoLaw2}
\begin{split}
d\hat{B}_{jn}(u,t) d\hat{B}^\dag_{j'n'}(u',t)&= dt \delta(u-u')\delta_{jj'}\delta_{nn'},\\
\end{split}
\end{equation}
with other entries in the It\^o table equal to zero. Finally, the explicit atom-bath coupling term is necessarily accompanied by the corresponding ``It\^o correction", given by the decay term $-(\Gamma_t\hat{\sigma}_{tt}+\Gamma_e\hat{\sigma}_{ee})/2$ in the first line of Eq.~\eqref{eq:QSSEfull}.

Having established the QSSE \eqref{eq:QSSEfull} as the basic dynamical equation for our model system, 
we will in the following subsection derive the  SME for the atom-cavity system. This SME describes the  evolution of the atom-cavity system under homodyne measurement of the cavity output, conditional to observing a particular homodyne current trajectory.

\subsection{Stochastic Master Equation for Homodyne Measurement}
\label{sec:SME_full}
Let us consider homodyne measurement of the output light of the cavity. In such a measurement, the output light from the cavity is mixed with a reference laser (a local oscillator), allowing the measurement of the quadrature $d\hat{Q}(t)$ of the 1D electromagnetic field bath~\cite{wiseman2009quantum,gardiner2015quantum},
\begin{equation}
\label{eq:quadrature}
d\hat{Q}(t) = d\hat{B}(t)e^{-i\phi} +d\hat{B}^\dagger(t) e^{i\phi},
\end{equation}
where $\phi$ is the phase of the local oscillator. The measurement will project the state of the bath onto an eigenstate of $d\hat{Q}(t)$ corresponding to the eigenvalue $dq(t)$, which defines the homodyne current via $dq(t)\equiv I(t)dt$. It can be shown~\cite{wiseman2009quantum,gardiner2015quantum} that the measurement outcome $dq(t)$ obeys a normal distribution  centered at the mean value of the cavity quadrature, i.e.,
\begin{equation}
\label{eq:eigenvalue_extfield1}
dq(t) \equiv I(t) dt = \sqrt{\kappa}\langle \hat{X}_{\phi}\rangle_c dt + dW(t),
\end{equation}
where $\hat{X}_{\phi}=e^{i\phi}\hat{c}^\dagger+e^{-i\phi}\hat{c}$ as defined in Sec.~\ref{sec2}, and $dW(t)$ is a random Wiener increment, which is related to the shot noise by $dW(t)= \xi(t)dt$. The expectation value $\langle\dots\rangle_c = {\rm Tr}(\dots\mu_c)$ is taken with a conditional density matrix $\mu_c$ of the joint atom-cavity system.

The evolution of $\mu_c$ is given by a SME derived from Eq.~\eqref{eq:QSSEfull} by projecting out the bath DOFs following  standard procedures~\cite{wiseman2009quantum,gardiner2015quantum},
\begin{align}
\label{completeSMEmaintext}
d\mu_c =&- \frac{i}{\hbar}[\hat{H}_{S},\mu_c ] dt+ \kappa\mathcal{D}[\hat{c}]\mu_c dt + \sqrt{\kappa}\mathcal{H}[\hat{c}e^{-i\phi}]\mu_c dW(t)\nonumber\\
&+\Gamma_t\!\int \!du N_{rt}(u) \mathcal{D}[e^{-i k_0 u \hat{z}}\hat{\sigma}_{rt}]\mu_cdt\nonumber\\
&- \frac{\Gamma_e P_{oe}}{2} \{ \hat{\sigma}_{ee},\mu_c\}dt\\
&+\Gamma_e\sum_{j=g,r}P_{je}\int du N_{je}(u) \mathcal{D}[e^{-i k_0 u \hat{z}}\hat{\sigma}_{je}]\mu_c\,dt .\nonumber
\end{align}
Here $\mathcal{D}[\hat{O}]\rho\equiv\hat{O}\rho\hat{O}^{\dagger}-\frac{1}{2}\hat{O}^{\dagger}\hat{O}\rho-\frac{1}{2}\rho\hat{O}^{\dagger}\hat{O}$ is the Lindblad superoperator, and ${\cal H}[\hat{O}]\rho\equiv\hat{O}\rho-{\rm Tr}(\hat{O}\rho)\rho+\text{H.c.}$ is a superoperator corresponding to homodyne measurement. As can be seen, the 1D electromagnetic field leads to both a decoherence term and a stochastic term [cf. the first line of Eq.~\eqref{completeSMEmaintext}] to the system evolution, which represent the backaction of homodyne measurement. In contrast, spontaneous emission into the 3D electromagnetic field is not continuously monitored in our model setup, and thus leads to pure decoherence, as captured by the last three lines of Eq.~\eqref{completeSMEmaintext}.

Incorporating the spontaneous emission terms in Eq.~\eqref{completeSMEmaintext} is important for a realistic description of an experiment. First, as is the case in most CQED experiments, we consider the levels $\ket{r}$ and $\ket{t}$ to form a closed cycling transition, such that $\ket{t}\to\ket{r}$ is the only dipole-allowed spontaneous emission channel for $\ket{t}$. Second, in contrast to $\ket{t}$, we allow $\ket{e}$ to have multiple spontaneous decay channels. This includes decays to states $\ket{g}$ and $\ket{r}$, with branching ratio $P_{ge}$ and $P_{re}$ respectively. Besides, $\ket{e}$ can also spontaneously decay outside the four level system, which is modeled as a pure decay term in the third line of Eq.~\eqref{completeSMEmaintext} with branching ratio $P_{oe}\equiv1-P_{ge}-P_{re}$. 

To summarize, the SME~\eqref{completeSMEmaintext} and the homodyne current~\eqref{eq:eigenvalue_extfield1} provides a complete description of the evolution of the joint atom-cavity system, subjected to continuous homodyne measurement of the cavity output in presence atomic spontaneous emission. In Eq.~\eqref{eq:eigenvalue_extfield1} and Eq.~\eqref{completeSMEmaintext}, the atomic DOFs includes both its external motion and its internal DOFs. In the next section, we will further reduce our equations to a model where only the cavity mode and atomic external motion appears, while we adiabatically eliminate the atomic internal dynamics assuming the atomic system remains in a dark state [compare Eq.~\eqref{eq: dark_state_def0}].

\subsection{Adiabatic Elimination of the Atomic Internal Dynamics}
\label{sec:elimination_internal}
As mentioned in Sec.~\ref{sec2}, we are interested in a regime where (i) the external motion of the atom is much slower than its internal dynamics, $|\hat{H}_{A,E}| \ll |\hat{H}_{A,I}| $, and (ii) the atom is coupled to the cavity mode dispersively, $\Delta_t\gg g(z)$. In this regime, according to Eq.~\eqref{completeSMEmaintext} we have a hierarchy of timescales, with the short time scale corresponding to the fast dynamics of the atomic internal DOFs, and the much longer time scale corresponding to the slow dynamics of cavity mode plus the atomic external DOFs.  This allows us to eliminate the atomic internal DOFs by an adiabatic assumption.

To be concrete, let us consider the eigenspectrum of $\hat{H}_{A,I}$ describing the atomic internal dynamics, which is shown in Fig.~\ref{fig:figureA1}. As mentioned in Sec.~\ref{sec2}, it includes a \emph{dark state},
\begin{equation}
\label{eq: dark_state_def}
\ket{D(z)}=\sin\alpha(z)\ket{g} -\cos\alpha(z)\ket{r},
\end{equation}
with the eigenenergy $E_D=0$. This state is spectrally well separated from the other eigenstates, namely the excited state $\ket{t}$, with the corresponding eigenenergy $E_{t}=-\hbar\Delta_t$; and the \emph{bright states}
\begin{equation}
\begin{split}
\label{eq: bright_state_def}
\ket{\pm} &=\frac{1}{\sqrt{2}}\Big[\pm\ket{e} +\cos\alpha(z) \ket{g}+\sin\alpha(z) \ket{r}\Big],
\end{split}
\end{equation}
with the corresponding energies $\smash{E_{\pm} (z)=\pm\hbar\Omega(z)/2}$. Here we have defined the total Rabi frequency $\smash{\Omega(z)\equiv[{\Omega_g^2 + \Omega_r^2(z)}]^{1/2}}$, and the mixing angle \begin{equation}
\label{eq:mixing angle 1}
\tan\alpha(z)=\Omega_r(z) / \Omega_g.
\end{equation}

In the absence of $\hat{H}_{AC}$ or $\hat{H}_{A,E}$, the internal state of the atom will remain in the dark state $\ket{D}$. Thus, the joint atom-cavity system is described by a product state
\begin{equation}
\label{eq:density_matrix_decomposition}
\mu_c (t) =\rho_c(t)\otimes\ket{D}\bra{D},
\end{equation}
where $\smash{\rho_c(t)\equiv{\rm Tr}_{A,I}[\mu_c(t)]}$ is the reduced density matrix for the cavity mode and the atomic external motion, with ${\rm Tr}_{A,I}$ indicating a trace over the atomic internal DOFs.

Reintroducing $\hat{H}_{AC}$ and $\hat{H}_{A,E}$ couples the dark state $\ket{D}$ to the rest of the atomic internal states. The Hamiltonian $\hat{H}_{AC}$ couples $\ket{D}$ and $\ket{t}$, see Eq.~\eqref{eq:JCcoupling}. The Hamiltonian $\hat{H}_{A,E}$ couples $\ket{D}$ and $\ket{\pm}$, due to the fact that the momentum operator $\hat{p}_z$ in $\hat{H}_{A,E}$ is non-diagonal in this position-dependent dark and bright states basis. Nevertheless, in the weak coupling limit $|\hat{H}_{A,E}|\ll \hbar\Omega(z)$, $|\hat{H}_{AC}|\ll \hbar|\Delta_t|$, the full density matrix $\rho_c(t)$ still preserves the separable form as in Eq.~\eqref{eq:density_matrix_decomposition}, except that for the atomic internal dynamics the dark state is weakly mixed with the excited states, which can be calculated in perturbation theory. The details of this derivation is presented in Appendix~\ref{app2}. The resulting evolution of the reduced density matrix $\rho_c(t)$ reads
\begin{align}
\label{eq:SME_firstElimination}
d\rho_c &=-\frac{i}{\hbar}\left[\hat{H}_{A,E} + \hat{H}_C + \hat{H}_{\rm coup} + V_{\rm na}(\hat{z}),\rho_c\right]dt \notag\\
&\quad+ \kappa \mathcal{D}[\hat{c}]\rho_{c}dt 
+ \sqrt{\kappa}\mathcal{H}[\hat{c}e^{-i\phi}]\rho_{c} dW(t)\nonumber\\
&\quad + \left(\frac{\Gamma_t}{\Delta_t^2}\mathcal{L}' + \mathcal{L}''\right)\rho_c dt.
\end{align}
Correspondingly, the homodyne current is determined according to
\begin{equation}
\label{eq:Ih_firstElimination}
dq(t) \equiv I(t) dt = \sqrt{\kappa}{\rm Tr}(\hat{X}_{\phi}\rho_c) dt + dW(t).
\end{equation}

In Eq.~\eqref{eq:SME_firstElimination}, $\hat{H}_{\rm coup}$ describes the local dispersive coupling [cf. Eq.~\eqref{eq:dispersive_coup}] between the atom and the cavity mode,
\begin{equation}
\label{eq:focusing Function}
\hat{H}_{\rm coup}=\frac{\hbar g^2(\hat{z})}{\Delta_t}\left[\cos\alpha(\hat{z})\right]^2\hat{c}^\dag\hat{c}\equiv \mathcal{A}f_{z_0}(\hat{z})\hat{c}^\dag\hat{c},
\end{equation}
and is parameterized in terms of a focusing function $f_{z_0}(\hat{z})$, see Eq.~\eqref{eq:dispersive_coup}, which is a dimensionless and normalized function around $z_0$, and a coupling strength $\mathcal{A}$, with the dimension of energy. 
We choose the normalization $\int dzf_{z_{0}}(z)=\ell_{0}$
with $\ell_{0}$ being the characteristic length scale of the system
under measurement, such that the matrix elements of $f_{z_{0}}(\hat{z})$ taken over the system states
are of order~$1$.
The spatial profile of $f_{z_0}(\hat{z})$ inherits from the position dependence of the mixing angle $\alpha(z)$, and can be engineered easily by adjusting the Rabi frequencies $\Omega_g$ and $\Omega_r(z)$ of the Raman lasers. In Sec.~\ref{sec:focusingFunction} we will discuss a laser configuration where $f_{z_0}(\hat{z})$ is peaked at $z_0$ with a nanoscale width.

Besides $\hat{H}_{\rm coup}$, Eq.~\eqref{eq:SME_firstElimination} contains several additional terms, which do not contribute to the homodyne measurement and thus are imperfections for the evolution of the system. This includes first of all 
\begin{equation}
\label{eq:Vna}
V_{\rm na}(z)=\frac{\hbar^2}{2m}(\partial_z\alpha)^2
\end{equation}
as the lowest order non-adiabatic potential for the atomic external motion~\cite{Lacki2006,Gorshkov2016}, originating from the spatially varying internal dark-state structure. As discussed in Sec.~\ref{sec:focusingFunction} below, we can make this term small with a proper choice of the laser configuration.

Note also that in a driven optical cavity, the atom cavity coupling
$\hat{H}_{{\rm coup}}$, Eq.~\eqref{eq:focusing Function}, gives
rise to an optical lattice potential $V_{{\rm OL}}(\hat{z})=\mathcal{A}f_{z_{0}}(\hat{z})\alpha_{0}^{2}$
for the atom with $\alpha_{0}=-2\mathcal{E}/\sqrt{\kappa}$ being
the amplitude of the stationary cavity field. This potential, however,
can be straightforwardly compensated by introducing a small detuning
$\Delta_{r}=g^{2}(z_{0})\alpha_{0}^{2}/\Delta_{t}$ for the $\ket{r}\to\ket{e}$
transition, resulting in an optical potential $V_{{\rm comp}}(\hat{z})=-\Delta_{r}\left[\cos\alpha(\hat{z})\right]^{2}$
for the atom, which compensates $V_{{\rm OL}}(\hat{z})$ in the spatial
region of interest.

The Liouvillian $(\Gamma_t/\Delta_t^2)\mathcal{L}'$ in the last line of (\ref{eq:SME_firstElimination}) describes the decoherence of the joint atom-cavity system, inherited from the atomic spontaneous emission in the channel $\ket{t}\to\ket{r}$, which is used to generate the dispersive atom-cavity coupling. The detailed expression of $\mathcal{L}'$ is given in Appendix~\ref{app2}. Here we note that, although this decoherence term can be suppressed by increasing the detuning $\Delta_t$, this will also reduce the dispersive coupling strength $\mathcal{A}$ [cf. Eq.~\eqref{eq:focusing Function}] and therefore the observed signal. As a consequence, $(\Gamma_t/\Delta_t^2)\mathcal{L}'$ serves as an intrinsic decoherence term for our dispersive atom-cavity coupling scheme, and will have important impact on the performance of the microscope, to be discussed in Sec.~\ref{sec4}.

Finally, $\mathcal{L}''$  in the last line of (\ref{eq:SME_firstElimination}) describes decoherence due to the coupling between $\ket{D}$ and $\ket{\pm}$ resulted from the motion of the atom, the detailed expression of which is given in Appendix~\ref{app2}. As shown there, $\mathcal{L}''$  can be made arbitrarily small by increasing the amplitude of $\Omega_g$ and $\Omega_r(z)$. We will neglect $\mathcal{L}''$ in the following.

To summarize, the SME  (\ref{eq:SME_firstElimination}) with internal states being eliminated, and the corresponding expression for the homodyne current (\ref{eq:Ih_firstElimination}) constitute the basic dynamical equations governing the time evolution of the quantum scanning microscope, and provide the basis of our discussion of the microscope operation in Sec.~\ref{sec4}.

\subsection{Engineering the Focusing Function $f_{z_0}(\hat{z})$}
\label{sec:focusingFunction}
\begin{figure}[t!]
\centering{}\includegraphics[width=0.49\textwidth]{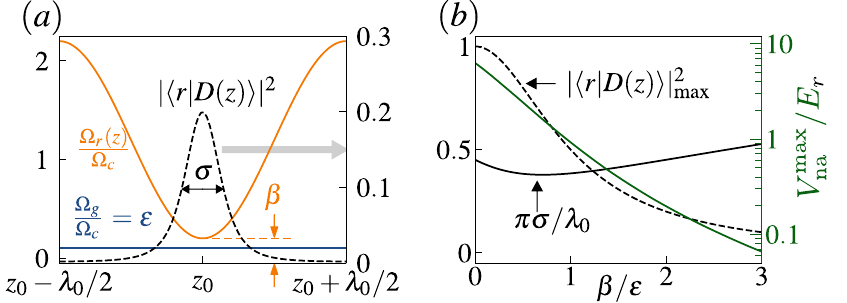} \caption{Engineering the focusing function $f_{z_0}(z)$, which enables subwavelength detection of the atomic density via Eq.~\eqref{eq:dispersive_coup}. (a) $f_{z_{0}}(z)\sim|\langle  r|D(z)\rangle|^2$ (dashed line) is shown together with the Rabi frequencies for the internal $\Lambda$-transition (see Fig.~\ref{fig:figure1}), $\Omega_r(z)=\Omega_{c}\{1+\beta-\cos[k_0(z-z_{0})]\}$ (light solid line) and $\Omega_g=\epsilon\Omega_{c}$ (dark solid line), for $\epsilon=\beta/2=0.1$. Adjusting $\beta$ and $\epsilon$ enables adjusting the resolution $\sigma$ (see text). (b) The resolution $\sigma$, the maximal overlap $|\langle r|D(z)\rangle|^2_{\rm max} $ and the maximal value $V_{\rm na}^{\rm max}$ of the non-adiabatic potential $V_{\rm na}(z)$ (in units of the recoil energy $E_r$) as a function of $\beta/\epsilon$, shown for $\epsilon=0.1$. For $\beta/\epsilon\gtrsim 1$, $V_{\rm na}(z)$ is strongly suppressed.}
\label{fig:figure3} 
\end{figure}
By eliminating the internal DOFs of the atom, we arrive at a local dispersive coupling between the atom and the cavity mode, Eq.~\eqref{eq:focusing Function}, defined via the focusing function $f_{z_0}(\hat{z})$. We now show how to engineer the resolution of the focusing function down to the deep subwavelength regime and with negligible non-adiabatic potential.

Let us consider the two Raman laser beams being phase coherent, with the Rabi frequencies parameterized by
\begin{equation}
\label{Omega1Expression}
\begin{split}
\Omega_g&=\epsilon \Omega_c,\\
\Omega_r(z)& = \Omega_c [ 1 + \beta - \cos k_0 (z-z_0) ],
\end{split}
\end{equation}
where $\Omega_c$ is a large reference frequency (assumed real positive for simplicity), and $\smash{0<\epsilon\sim\beta\ll1}$. In practice, $\Omega_r(z)$ can be realized, e.g., by super-imposing phase-coherent laser beams to form the standing wave $\Omega_c\cos k_0(z-z_0)$ along the $z$-axis, and another standing wave in an orthogonal direction, to provide the offset $\Omega_c(1+\beta)$, as shown in the Fig.~\ref{fig:figure1}(d).

The laser configuration \eqref{Omega1Expression} completely determines the focusing function, cf. Eqs.~\eqref{eq:mixing angle 1} and \eqref{eq:focusing Function}, which is shown in Fig~\ref{fig:figure3}(a). We define the resolution $\sigma$ of the focusing function as its full width at half maximum (FWHM) which reads
\begin{equation}
\label{appResolution}
\frac{\sigma}{\lambda_0} = \frac{\sqrt{2\beta}}{\pi}\left[\sqrt{2 + (\epsilon/\beta)^2}-1\right]^{1/2},
\end{equation}
with $\lambda_0=2\pi/k_0$ being the wavelength for the $\ket{r}\to\ket{e}$ transition. Thus, the resolution can be made subwavelength, by choosing $\epsilon\sim\beta\ll 1$.

Moreover, the laser configuration \eqref{Omega1Expression} allows for rendering the non-adiabatic potential $V_{\rm na}(z)$ negligible [cf. Eq.~\eqref{eq:Vna}]. In the regime of interest $\epsilon\sim\beta\ll 1$, such a non-adiabatic potential expanded around focusing point $z_0$ can be calculated with Eqs.~\eqref{eq:mixing angle 1} and~\eqref{eq:Vna} as
\begin{equation}
\label{nonAdiabaticCorrection}
V_{\rm na}(z)=E_r\left\{\frac{4\epsilon k_0 (z-z_0)}{[k^2_0(z-z_0)^2+2\beta]^2+4\epsilon^2}\right\}^2,
\end{equation}
where $E_r=\hbar^2k_0^2/(2m)$ is the recoil energy of the atom. As illustrated in Fig.~\ref{fig:figure3}(b), $V_{\rm na}(z)$ decreases rapidly with increasing the ratio $\beta/\epsilon$. 
Physically,
decreasing $\epsilon/\beta$ reduces the overlap between the dark
state and the state $\ket{r}$, $|\langle r|D(z)\rangle|_{{\rm max}}^{2}=(1+\beta^{2}/\epsilon^{2})^{-1}$,
such that the dark state varies more slowly in space, thus suppressing
the corresponding non-adiabatic potential. 

In fact, in the considered limit $\sigma\ll\lambda_{0}$, one can
make both $\sigma$ and $V_{{\rm na}}(z)$ arbitrary small by choosing
appropriate values of $\beta$ and $\epsilon/\beta$ which scale as
$\beta\sim(\sigma k_{0})^{2}$ and $\epsilon/\beta\sim\sigma k_{0}\sqrt{V_{{\rm na}}^{\mathrm{max}}/E_{r}}$,
where $V_{{\rm na}}^{\mathrm{max}}=\max_{z}\{V_{{\rm na}}(z)\}$ is
the maximal value of the nonadiabatic potential $V_{{\rm na}}(z)$.
As a consequence, (i)~the microscope resolution is unlimited at the
level of the focusing function engineering and (ii)~the non-adiabatic
potential $V_{{\rm na}}(\hat{z})$ can be neglected in the SME~\eqref{eq:SME_firstElimination}
hereafter. However, with decreasing $\epsilon/\beta$ one also reduces
the signal strength in the photocurrent which is proportional to the
population of the $\ket{r}$ state, such that a longer measurement time
is required to distinguish it from the shot noise. The long measurement
time in turn increases the role of spontaneous emission processes
which ultimately limit the microscope resolution, as discussed below
in Sec.~\ref{sec:resolution_limit}.

We comment that, although we have focused here on a standing-wave implementation of the focusing function, there exist alternative 
designs of the laser profiles for dark-state microscopy, e.g., exploiting optical vortices created by holographic techniques or by Laguerre-Gaussian beams~\cite{Gorshkov2008,Maurer2010}. 

\subsection{Filtered Homodyne Current and the SNR}
\label{sec: sigfiltering}
In this section we discuss the  signal filtering for homodyne detection, and thus define the signal-to-noise ratio (SNR), which will serve as a key figure of merit to quantify the performance of the microscope below.

The homodyne current [see e.g., Eq.~\eqref{eq:homodyne_review} or \eqref{eq:Ih_firstElimination}] is noisy, as it contains the (white) shot noise corresponding to the Wiener increment $dW(t)$, inherited from the vacuum fluctuation of the electromagnetic bath. The shot noise can be suppressed by filtering the homodyne current with a linear lowpass filter,
\begin{equation}
\mathcal{I}_{\tau}(t)=\int  dt'h_{\tau}(t-t')I(t').
\label{eq:filteredIh}
\end{equation}
Here, $\mathcal{I}_{\tau}(t)$ is the \emph{filtered homodyne current}, while $h_{\tau}(t)$ is the filter function with a frequency bandwidth characterized by $1/\tau$. The filter passes the low frequency components of the homodyne current including the conditional expectation value $\sqrt{\kappa} \langle\hat{X}_\phi\rangle_c$, and attenuates the high frequency component of the shot noise, reducing its variance to $\sim1/\tau$. With diminished shot noise, the filtered homodyne current allows us to directly read out the signal which, as will be discussed in detail in Sec.~\ref{sec4}, maps out the spatial density of the atomic system. 

The quality of the filtered homodyne current is reflected by its signal to noise ratio (SNR), i.e., the relative power between the signal and the noise, defined as
\begin{equation}
\label{eq:SNRdef}
{\rm SNR}(t) = \frac{\langle{\cal I}_{\tau}(t)\rangle_{{\rm st}}^{2}}{\langle{\cal I}_{\tau}^{2}(t)\rangle_{{\rm st}}-\langle{\cal I}_{\tau}(t)\rangle_{{\rm st}}^{2}},
\end{equation}
where $\langle\dots\rangle_{\rm st}$ denotes statistical averaging over all measurement runs. We note that on the RHS of Eq.~\eqref{eq:SNRdef} the total noise variance (in the denominator) includes not only the filtered shot noise, but also the noise inherited from the fluctuating signal $\sqrt{\kappa} \langle \hat{X}_{\phi}\rangle_c$, cf. Eqs.~\eqref{eq:homodyne_review} and~\eqref{eq:filteredIh}. This noise component is a manifestation of the measurement backaction.  

Both the filtered homodyne current $\mathcal{I}_{\tau}(t)$ and its SNR depend on the choice of the filter function $h_\tau(t)$, in particular its inverse bandwidth $\tau$. On the one hand, $\tau$ should be chosen as large as possible, to suppress the shot noise thus to enhance the SNR; on the other hand, $\tau$ should be kept small enough for the signal to pass through. The optimal $\tau$ will depend on the systems under the observation, and the operation modes of the microscope. This will be discussed in detail in Sec.~\ref{sec4}. In contrast to the bandwidth, the exact shape of $h_\tau(t)$ has much smaller impact on both the filtered homodyne current and its SNR, and a simple filter suffices to illustrate the main features of the measured quantity. In this paper, we adopt the filter
\begin{equation}
\label{simpleFilter}
h(t)=\left\{ \begin{array}{ll}
\tau^{-1} e^{- t/\tau},\qquad &t\geq 0, \\
0,\qquad &t<0.
\end{array}\right.
\end{equation}
Besides its simplicity, it has the additional benefit that the corresponding SNR Eq.~\eqref{eq:SNRdef} can be calculated efficiently with a numerical method as detailed in Appendix~\ref{sec:SNRcascade}. 

\subsection{Summary of the Model}

The key result of the present section is the SME Eq.~\eqref{eq:SME_firstElimination} describing the dynamics of the quantum scanning microscope, together with the corresponding homodyne current (i.e., the measurement signal) Eq.~\eqref{eq:Ih_firstElimination}. 
%Eq.~\eqref{eq:SME_firstElimination} features a dispersive atom-cavity coupling $\hat{H}_{\rm coup}$ defined in terms of the \emph{subwavelength} focusing function $f_{z_0}(\hat{z})$, cf. Eq.~\eqref{eq:focusing Function}, which can be easily manipulated by adjusting the laser intensities as discussed in Sec.~\ref{sec:focusingFunction}. The homodyne current Eq.~\eqref{eq:Ih_firstElimination}, after filtering, allows us to directly read out the information of the atom-cavity system, and its quality is quantified in terms of the SNR Eq.~\eqref{eq:SNRdef}. The SME Eq.~\eqref{eq:SME_firstElimination} further includes the decoherence inherited from the atomic spontaneous emission, captured dominantly by $\mathcal{L}'$, as the imperfection due to the dispersive coupling scheme used by the microscope. 
Altogether, they allow us to study the various operation modes of the microscope and examine its performance in the presence of the atomic spontaneous emission, to be detailed in Sec.~\ref{sec4}. 

\section{Microscope Operation}
\label{sec4}
With the quantum optical model at hand, we discuss below in detail the operation of the microscope.
The microscope is characterized by three parameters: the spatial resolution $\sigma\ll\lambda_0$, the temporal resolution $\tau_{c}=1/\kappa$ with $\kappa$ the cavity linewidth, and the dispersive atom-cavity coupling $\mathcal{A}$ controlling the measurement strength [see Eq.~\eqref{eq:focusing Function}]. As we will show, the bad (good) cavity limit, defined as the cavity linewidth $\kappa$ being much larger (smaller) than the frequency scale of atomic motion, corresponds to two operating modes of the microscope --- which we call the {\em Movie Mode} and {\em Scanning Mode}, respectively. These operation modes feature distinct effective observables of the atomic system, thus providing different strategies for measuring the atomic density, with different applications.

In the following we explore these operating modes, by analyzing the observables being measured, and the corresponding measurement backaction. We illustrate these features via numerical simulation of the measurement for a simple example system, an atom moving in a harmonic oscillator (HO) potential, $V(z)=m\omega^2z^2/2$, with the atomic mass $m$, trap frequency $\omega$ and motional eigenstates $|n\rangle$, $n=0,1,2\dots$ For this system, the {Movie Mode} ({Scanning Mode}) is defined by $\kappa\gg\omega$ ($\kappa\ll\omega$) respectively. Further, to resolve the spatial density distributions, we require a spatial resolution better than the length scale $\ell_0\equiv\sqrt{\hbar/m\omega}$ set by the HO ground state, $\sigma\lesssim \ell_0$. %For the {Movie Mode}, we demonstrate its ability to monitor time-resolved density distributions, by preparing the HO in a coherent state and measuring its oscillation across a fixed focal point. For the {Scanning Mode}, we demonstrate its power to prepare and to scan energy eigenstates of the HO in a single measurement run with high SNR. 
We include in these discussions decoherence due to spontaneous emission as imperfection, aiming at providing a direct reference for experimental implementations of the microscope.

\subsection{Bad Cavity Limit: the Movie Mode of the Microscope}
\label{badCavityLimit}
In the bad cavity limit the cavity dynamics is much faster than the atomic motion such that the former instantaneously follows the latter. Such a time scale separation allows us to adiabatically eliminate the cavity mode in Eqs.~\eqref{eq:SME_firstElimination} and~\eqref{eq:Ih_firstElimination}, resulting in an equation for the atomic density matrix $\tilde{\rho}_c\equiv{\rm Tr}_C(\rho_c)$ alone, where ${\rm Tr}_C$ indicates a trace over the cavity mode. Carrying out this elimination (see Appendix~\ref{adiabaticEliminationCavity}) results in the following expression for the homodyne current [see also Eq.~\eqref{eq:Homodyne_movie}]:
\begin{equation}
\label{eq:homodyne_elimination_badcavity}
dq(t) \equiv I(t)dt = 2\sqrt{\gamma}\langle f_{z_{0}} (\hat{z})\rangle_{c}\,dt + dW(t),
\end{equation}
i.e. the homodyne current directly reflects the expectation value of the focusing function $f_{z_{0}}(\hat{z})$. In Eq.~\eqref{eq:homodyne_elimination_badcavity}
\begin{equation}
\label{eq:gamma_definition}
\gamma=\left(\frac{4\mathcal{A}\mathcal{E}}{\hbar\kappa}\right)^{2}
\approx \left(4\mathcal{EC}\frac{\sigma}{\ell_{0}}\frac{\Gamma_t}{\Delta_t}|\langle r|D(z)\rangle|^2_{\rm max}\right)^{2}
\end{equation}
is the measurement rate with $\mathcal{C}=g^2(z_0)/\kappa\Gamma_t$ the cavity cooperativity, and we have chosen the homodyne angle as $\phi=-\pi/2$ to maximize the homodyne current (see Appendix~\ref{adiabaticEliminationCavity}). Correspondingly, the evolution of the atomic system is given by
\begin{align}
\label{eq:SME_elimination_badcavity}
d\tilde{\rho}_{c} & =-\frac{i}{\hbar}[\hat{H}_{{A,E}},\tilde{\rho}_{c}]\,dt + \gamma{\cal D}[f_{z_{0}}(\hat{z})]\tilde{\rho}_{c}\,dt\nonumber \\
 & \quad+\sqrt{\gamma}{\cal H}[f_{z_{0}}(\hat{z})]\tilde{\rho}_{c}\,dW(t)
 + \frac{\gamma}{4\mathcal{C}}\mathcal{L}_{\rm sp}\tilde{\rho}_c\,dt.
\end{align}
Here the first term on the RHS describes the coherent evolution of the atom according to its motional Hamiltonian $\hat{H}_{{A,E}}$, while the second and the third terms describe the backaction resulting from continuous measurement of $f_{z_0}(\hat{z})$. The last term corresponds to decoherence of the motional density matrix of the atom due to spontaneous emission, 
\begin{equation}
\label{eq:Lsp_maintext}
\begin{split}
&\mathcal{L}_{{\rm sp}}\tilde{\rho}_{c} =\vphantom{\int}\mathcal{D}\left[f_{z_{0}}(\hat{z})\right]\tilde{\rho}_{c}+P_{re}\mathcal{D}\left[f_{z_{0}}(\hat{z})\sin\hat{\alpha}_{z_0}\right]\tilde{\rho}_{c}\\
&+P_{ge}\mathcal{D}\left[f_{z_{0}}(\hat{z})\tan\hat{\alpha}_{z_0}\sin\hat{\alpha}_{z_0}\right]\tilde{\rho}_{c}\\
&-\frac{1}{2}\left\{ f_{z_{0}}^{2}(\hat{z})[\tan^{2}\hat{\alpha}_{z_0}(1-P_{ge}\sin^{2}\hat{\alpha}_{z_0})-P_{re}\sin^{2}\hat{\alpha}_{z_0}],\tilde{\rho}_{c}\right\},
\end{split}
\end{equation}
which will be analyzed in detail in Sec.~\ref{sec:Lsp}.

Eqs.~\eqref{eq:homodyne_elimination_badcavity} and~\eqref{eq:SME_elimination_badcavity} provide a complete description of the quantum evolution of the atom subjected to continuous measurement in the {Movie Mode} of the microscope. In the following we study the measurement and its backaction as well as effects of spontaneous emission, which we illustrate with the example of monitoring wave packet dynamics in a harmonic oscillator potential.

\subsubsection{Observable and the measurement backaction}
\label{sec:observable_bad_cavity}
\begin{figure}[t!]
\centering{}\includegraphics[width=0.49\textwidth]{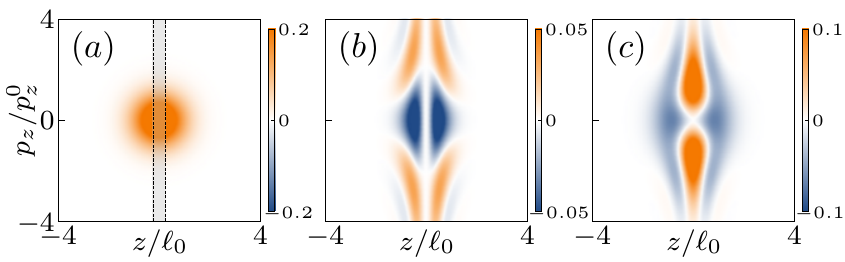} 
\caption{Visualization of the effect of the measurement backaction in the {Movie Mode} of the microscope, via the Wigner function of (a) $\tilde{\rho}_{c}=\ket{0}\bra{0}$, (b) $\mathcal{D}[f_{z_0}(\hat{z})]\tilde{\rho}_{c}$ and (c) $\mathcal{H}[f_{z_0}(\hat{z})]\tilde{\rho}_{c}$, for an atom in a HO potential. The focal region is shown as the dashed area in (a), with focal point $z_0=0$ and resolution $\sigma=0.5\ell_0$, where $\ell_0\equiv\sqrt{\hbar/m\omega}$ is the HO length scale. Comparing (a) and (b), we see that the decoherence term $\mathcal{D}[f_{z_0}(\hat{z})]\tilde{\rho}_{\rm th}$ increases the momentum fluctuation of the atom. Comparing (a) and (c), we find that $\mathcal{H}[f_{z_0}(\hat{z})]\tilde{\rho}_{\rm th}$ induces a redistribution of the population between the focal region and the outside (see text).}
\label{fig:figure4} 
\end{figure}

The observable in the Movie Mode is the focusing function $f_{z_0}(\hat{z})$, cf.~Eqs.~\eqref{eq:homodyne_elimination_badcavity}, which provides information of the local atomic density with a resolution $\sigma$ [cf. Eq.~\eqref{appResolution}], where in the limit $\sigma \rightarrow 0$,  $f_{z_0}(\hat{z}) \sim \delta(\hat  z - z_0) = \ket{z_0}\bra{z_0}$. Given that $f_{z_0}(\hat{z})$ does not not commute with the Hamiltonian of the atomic system, $[\hat{H}_{A,E}, f_{z_0}(\hat{z})]\neq 0$, the Movie Mode is obviously not QND, i.e.~measurement backaction competes with the coherent dynamics generated by $\hat{H}_{A,E}$.

The measurement backaction is represented by terms ${\cal D}[f_{z_{0}}(\hat{z})]\tilde{\rho}_{c}$ and ${\cal H}[f_{z_{0}}(\hat{z})]\tilde{\rho}_{c}$ in Eq.~\eqref{eq:SME_elimination_badcavity}. To visualize the action of these terms, we plot in Fig.~\ref{fig:figure4}  the corresponding Wigner functions in phase space, where we take the ground motional state $\tilde{\rho}_{c}=\ket{0}\bra{0}$ of the atom in the HO potential as the reference state [with Wigner function plotted in Fig.~\ref{fig:figure4} (a)].
As can be seen in Figs.~\ref{fig:figure4}(b) and (c), respectively, the decoherence term $\mathcal{D}[f_{z_0}(\hat{z})]$ induces momentum diffusion of the atom, while the homodyne term $\mathcal{H}[f_{z_0}(\hat{z})]$ continuously projects the atomic state inside/outside the focal region, by stimulating population flow between these two regions. Both terms elongate the Wigner function of the atom along  $p_z$, manifesting the enhanced momentum fluctuation due to the measurement.

\begin{figure}[t!]
\centering{}\includegraphics[width=0.49\textwidth]{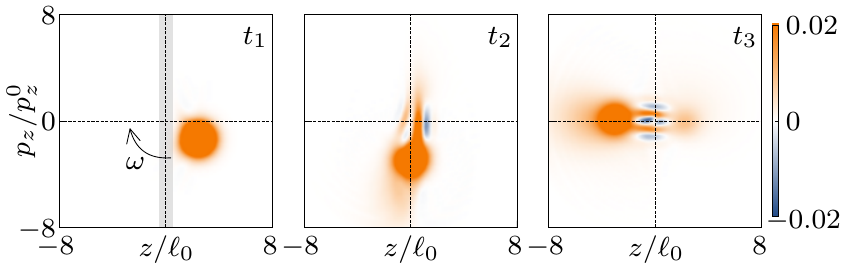} \caption{Dynamics of the measurement in the {Movie Mode} of the microscope. We show the Wigner function of a coherent state $\ket{\alpha}$, evolving in time according to Eq.~\eqref{eq:SME_elimination_badcavity}, and average over all measurement outcomes to yield a deterministic evolution. At $t_1=0.1T_{\rm osc}$ ($T_{\rm osc}\equiv2\pi/\omega$) the coherent state has not yet entered the focal region (shown as the grey area), and remains an equal distribution of position and momentum uncertainty. At $t_2=0.25T_{\rm osc}$, as the coherent state crosses into the focal region (centered at $z_0=0$), the distribution begins to spread out along the $p_z$-axis. At $t_3=0.5T_{\rm osc}$, the increased uncertainty is fed into the $z$-axis, increasing the fluctuation of the measured homodyne current (see text). Parameters are chosen as $\alpha=2$, $\sigma=0.5\ell_{0}$ and $\gamma=4\omega$. }
\label{fig:figure5} 
\end{figure}

To visualize the competition between the measurement backaction and the Hamiltonian,  we show in Fig.~\ref{fig:figure5} for different times the Wigner function of an atom in a coherent state evolving according to Eq.~\eqref{eq:SME_elimination_badcavity}, and averaged over all measurement outcomes (i.e. as solution of the quantum master equation), where for simplicity we set $\mathcal{L}_{\rm sp}=0$ in Eq.~\eqref{eq:SME_elimination_badcavity}. As the atom passes through the focal region (here fixed at $z_0=0$), its momentum fluctuation is increased due to the measurement backaction, causing its Wigner function to spread out along the $p_z$-axis (cf. Fig.~\ref{fig:figure5} at time $t_2$). At later times, the Wigner function, now including the spread in momentum, continues to rotate with frequency $\omega$, and thus leading to extra fluctuation along $z$ (cf. Fig.~\ref{fig:figure5} at time $t_3$). The pattern appearing near the focal region in Fig.~\ref{fig:figure5}(b-c) results from the coherent interference between the transmitted component and the reflected component of the atomic wavepacket when crossing the ``dissipative barrier'' in the focal region.

\subsubsection{Decoherence due to spontaneous emission}
\label{sec:Lsp}
Spontaneous emission induces extra decoherence in atomic motion, captured by Eq.~\eqref{eq:Lsp_maintext}. This term is derived from $\mathcal{L}'$ of Eq.~\eqref{eq:SME_firstElimination} [given in Eq.~\eqref{SPsuperOP1}]. The first two lines of Eq.~\eqref{eq:Lsp_maintext} describe the momentum diffusion of the atomic wave packet in the focal region during the excitation--spontaneous-emission cycle involving the states $\ket{t}$ and $\ket{r}$. 
The third line of Eq.~\eqref{eq:Lsp_maintext} describes the gradual loss of atoms due to the decay from the bright states $\ket{\pm}$ to levels lying outside the four level system under consideration (see Fig.~\ref{fig:figureA1}), which makes atoms invisible to the microscope. Altogether, these decoherence processes are strongly suppressed for cavities with high cooperativity, since their rate is given by~$\gamma/4\mathcal{C}$ [cf. Eq.~\eqref{eq:SME_elimination_badcavity}].

To quantitatively understand the role of these extra decoherence processes, below we perform numerical simulation of the evolution of the atomic system subjected to continuous measurement and spontaneous emission, where the measurement rate $\gamma$ and cooperativity $\mathcal{C}$ are chosen according to realistic experimental parameters (given in Sec.~\ref{sec:exp_considerations}). The simulations confirm that the atomic spontaneous emission brings in negligible detriment to the performance of the microscope for $\mathcal{C}\gg 1$, which can be achieved in state-of-the-art cavity QED experiments~\cite{Barontini2015}.

\subsubsection{Application: monitoring wave packet oscillations}
\label{simulation_bad_cavity}
We demonstrate the Movie Mode of the microscope, as monitoring the oscillation of an atomic wave packet released in a HO potential. As shown below, the bad cavity condition $\kappa\gg\omega$ provides the \emph{time resolution} to `take a movie' of time-dependent density distributions. This is accompanied, due to the non-QND nature of the measurement, by the competition between Hamiltonian dynamics and measurement backaction, and as a consequence, by  limitations due to achievable SNR in a single measurement run.

\begin{figure}[t!]
\centering{}\includegraphics[width=0.49\textwidth]{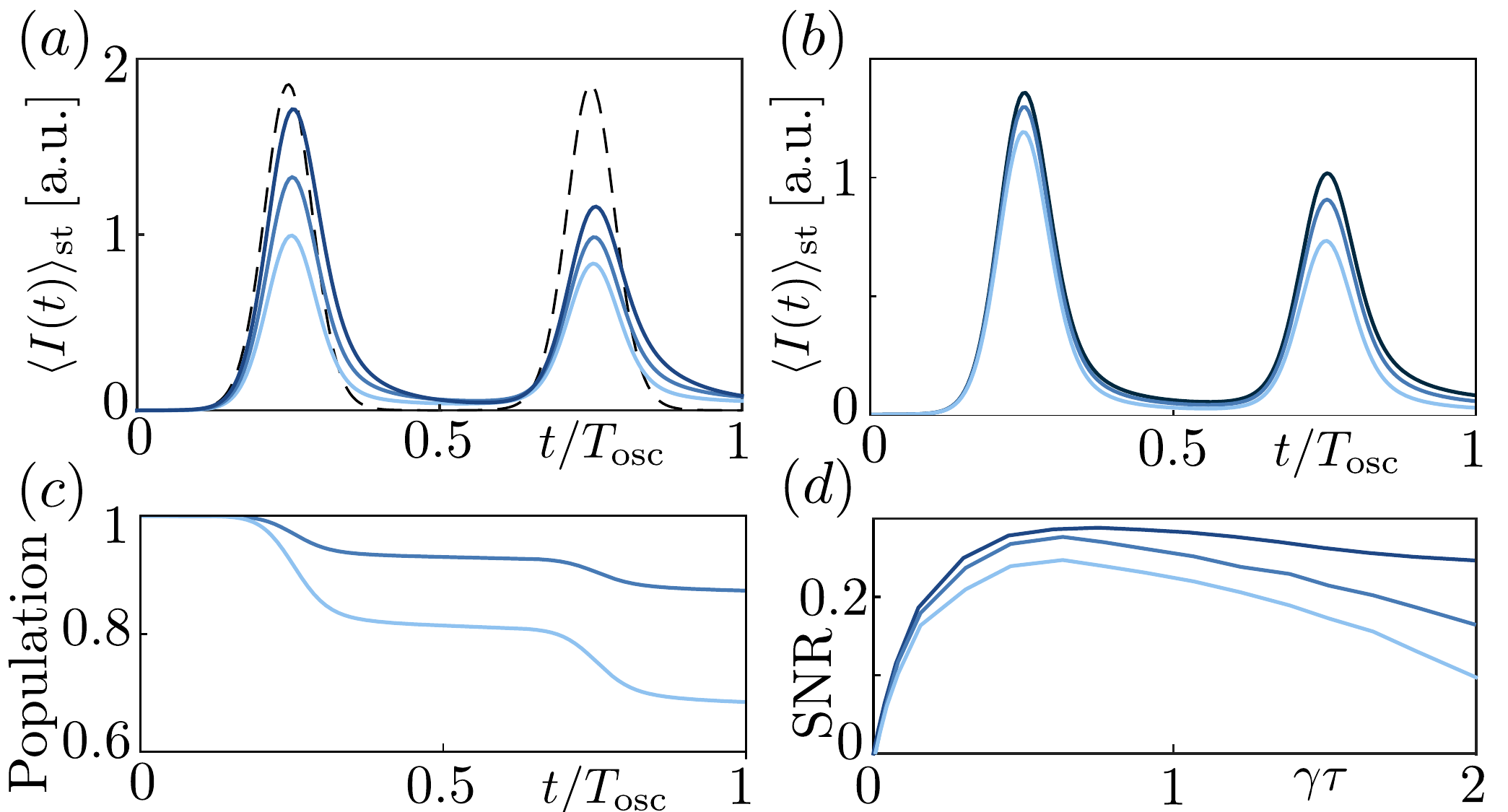} \caption{Monitoring wave packet oscillations in the {Movie Mode} of the microscope. (a) The statistically-averaged homodyne current $\langle I(t)\rangle_{\rm st}$ over the oscillation period $T_{\rm osc}\equiv2\pi/\omega$ with increasing measurement rate $\gamma=1,2,4\omega$ (light to dark) and without spontaneous emission (corresponding to $\mathcal{C}=\infty$). Dashed line indicates the ideal transit signal with no measurement backaction (arbitrary unit). (b) $\langle I(t)\rangle_{\rm st}$ for $\gamma=2\omega$, and $\mathcal{C}=3,10,\infty$ (light to dark). Lower $\mathcal{C}$ corresponds effectively to stronger spontaneous emission, which degrades the homodyne current. (c) decay of the dark state population due to spontaneous emission, with the same parameters as in (b). (d) SNR at the
first peak ($t=0.25T_{{\rm osc}}$) for a single measurement run, as a function of $\gamma\tau$ with $\tau$ being the filter integration time (see text), for $\mathcal{C}=3,10,\infty$ (light to dark).}
\label{fig:figure6} 
\end{figure}

We initialize the atom in a coherent state $\ket{\alpha}$ and monitor its subsequent oscillations by focusing the microscope at the trap center $z_0=0$. Without coupling to the microscope, the atomic wave packet will pass through the trap center with a velocity $v=\sqrt{2}\ell_0 |\alpha|\omega$ at times $t=1/4T_{\rm osc}, 3/4T_{\rm osc}$ etc., where $T_{\rm osc}\equiv2\pi/\omega$ is the oscillation period. Once coupled to the microscope, the atom will evolve according to Eq.~\eqref{eq:SME_elimination_badcavity}, characterized by a competition between free oscillation and measurement backaction. Such a competition is necessarily reflected in the measurement records of the microscope. To show this, in Fig.~\ref{fig:figure6}(a) we plot for different measurement strengths $\gamma$ the ensemble-averaged homodyne current $\langle I(t)\rangle_{\rm st}$, with $I(t)$ given by Eq.~\eqref{eq:homodyne_elimination_badcavity} and $\langle\dots\rangle_{\rm st}$ standing for statistical average. As can be seen clearly, $\langle I(t)\rangle_{\rm st}$ represents faithfully the shape of the atomic wave packet passing through the focusing region, and reflects the measurement backaction as a successive distortion of the signal with time, which becomes more significant with increasing $\gamma$. The impact of atomic spontaneous emission is illustrated in Fig.~\ref{fig:figure6}(b), where we plot $\langle I(t)\rangle_{\rm st}$ for different strengths of spontaneous emission $\gamma/4\cal{C}$ at a fixed measurement strength $\gamma$.
As shown there, spontaneous emission diminishes the measured homodyne current --- the smaller $\mathcal{C}$, the weaker the homodyne current. This is a direct consequence of the gradual depletion of the population in the internal dark state due to spontaneous emission, shown in Fig.~\ref{fig:figure6}(c).

The fact that the measurement backaction competes with the Hamiltonian evolution provides a limitation on the achievable SNR of the filtered homodyne current in a single measurement run. To illustrate this, we plot  in Fig.~\ref{fig:figure6}(d)  the SNR of a single measurement run against the dimensionless measurement strength $\gamma\tau$, where $\tau$ is the filter integration time (cf. Sec.~\ref{sec: sigfiltering}). We choose an `optimal' $\tau$ defined via $\tau = \sigma / v$, with $\sigma$ the microscope resolution and $v$ the group velocity of the atomic wave packet at $z=0$. It allows the signal (which has a bandwidth $\sim v/\sigma$ due to the finite resolution $\sigma$ of the focusing function) to pass through, while filtering out the shot noise with frequencies outside the defined bandwidth. In Fig.~\ref{fig:figure6}(d), at small measurement strength $\gamma$, the SNR grows with $\gamma$, due to the enhancement of the signal relative to the shot noise. At large $\gamma$, SNR eventually drops down because of the strong measurement backaction. The appearance of such an upper bound of the SNR in a single measurement run is a general feature of non-QND measurements~\cite{Clerk2010}. Fig~\ref{fig:figure6}(d) also includes the effect of the atomic spontaneous emission. As expected, spontaneous emission reduces the SNR further, since it diminishes the measured signal [cf. Fig.~\ref{fig:figure6}(b)].

\subsection{Good Cavity Limit: the Scanning Mode of the Microscope}
\label{sec3}
We now consider the good cavity limit $\kappa \ll \omega$ where, as mentioned in Sec.~\ref{sec:operationPrinciples}, the cavity effectively filters out fast dynamics of the atomic system. This can be seen  by transforming the SME (\ref{eq:SME_firstElimination}) to an interaction picture with respect to $\hat{H}_{A,E}$, the motional Hamiltonian of the atom. In this picture the focusing function $f_{z_0}(\hat{z})$ becomes time-dependent and can be expanded as $\sum_{\ell}\hat f_{z_{0}}^{(\ell)}e^{-i\ell\omega t}$, with the $\ell$-th \emph{sideband component} $\hat f_{z_{0}}^{(\ell)}\equiv\sum_{n}f_{n,n+\ell}|n\rangle\langle n+\ell|$, and $f_{mn}\equiv\langle m|f_{z_{0}}(\hat{z})|n\rangle$. Due to the narrow cavity linewidth $\kappa\ll \omega$, the cavity will effectively enhance the light coupling to the zero frequency component $\hat{f}^{(0)}_{z_0}$ by averaging out the rest $\hat{f}^{(\ell)}_{z_0}$ with $\ell\neq 0$. Consequently, only $\hat{f}^{(0)}_{z_0}$ is reflected in the homodyne current. As will be detailed in Sec.~\ref{sec:eQND}, the operator $\hat{f}^{(0)}_{z_0}$ serves as an \emph{emergent QND} (eQND) observable of the atomic density, which allows for mapping out the spatial density of \emph{energy eigenstates} of the atom in a single measurement run with a high SNR.

In the good cavity limit, we can again eliminate the cavity mode to obtain equations of motion for the atomic system alone, the detailed derivation of which is presented in Appendix~\ref{adiabaticEliminationCavity}. Summarizing the results, the expression for the homodyne current is given by [see also Eq.~\eqref{eq:Homodyne_Scan}]
\begin{equation}
\label{eq:homodyne_good_cavity_maintext}
dq(t) = I(t)dt = 2\sqrt{\gamma}\,\langle \hat f_{z_{0}}^{(0)}\rangle_{c}\,dt + dW(t).
\end{equation}
We see, as mentioned above, the homodyne current following the expectation value of the eQND variable $\hat{f}^{(0)}_{z_0}$. Correspondingly, the evolution of the atomic system becomes
\begin{align}
d\tilde{\rho}_{c} & =-\frac{i}{\hbar}[\hat{H}_{{A,E}},\tilde{\rho}_{c}]\,dt+\sum_{\ell\neq0}\frac{\gamma}{1+(2\omega\ell/\kappa)^{2}}{\cal D}[\hat f_{z_{0}}^{(\ell)}]\tilde{\rho}_{c}\,dt\nonumber \\
 & \quad+\gamma{\cal D}[\hat f_{z_{0}}^{(0)}]\tilde{\rho}_{c}\,dt+\sqrt{\gamma}{\cal H}[\hat f_{z_{0}}^{(0)}]\tilde{\rho}_{c}\,dW(t)\notag\\
 &\quad+\frac{\gamma}{4\mathcal{C}}\mathcal{L}_{\rm sp}\tilde{\rho}_cdt.\label{eq:SME_good_cavity}
\end{align}
Here $\gamma$ is the measurement strength defined by Eq.~\eqref{eq:gamma_definition}, and $\mathcal{L}_{\rm sp}$ describes the extra decoherence introduced by atomic spontaneous emission, cf. Eq.~\eqref{eq:Lsp_maintext}. In Eq.~\eqref{eq:SME_good_cavity}, the superoperator for homodyne measurement contains only $\hat{f}_{z_0}^{(0)}$, while higher sideband components $\hat{f}_{z_0}^{(\ell)}$ with $\ell\neq 0$ only induce decoherence captured by the Lindblad operators $\mathcal{D}[\hat{f}_{z_0}^{(\ell)}]$, with a diminished rate $\gamma[1+(2\omega\ell/\kappa)^2]^{-1}\to 0$ as suppressed by the cavity in the limit $\kappa/\omega\to0$. This fact causes distinct measurement backaction to the atomic system compared to the bad cavity limit in Sec.~\ref{badCavityLimit}, as will be discussed in detail below. 

\subsubsection{Emergent QND measurement of atomic density}
\label{sec:eQND}
\begin{figure}[t!]
\centering{}\includegraphics[width=0.49\textwidth]{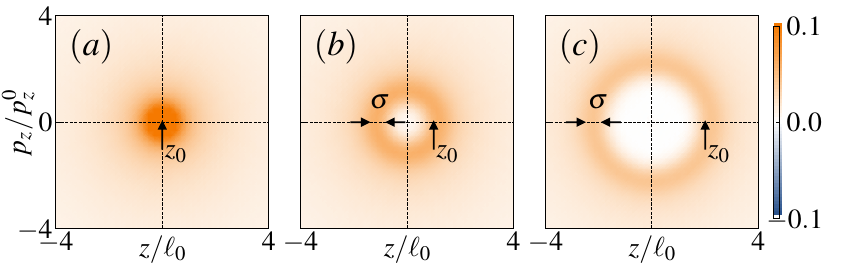} \caption{The eQND observable $\hat{f}_{z_0}^{(0)}$ in the {Scanning Mode} of the microscope. In (a-c), we show the Wigner function of $\hat{f}_{z_0}^{(0)}$ for three different focal points $z_0=0,\ell_0,{2\ell_0}$ respectively, with $\ell_0$ the HO length scale. In (b) and (c), the Wigner function manifests as a doughnut centered on the origin, of radius $z_0$ and of width $\sigma$. Resolution $\sigma=0.5\ell_0$.}
\label{fig:figure7} 
\end{figure}

In Ref.~\cite{Yang2017a}, we introduced the concept of eQND measurement, applicable to an arbitrary observable $\hat{\cal{O}}$ which, in general, does not necessarily commute with the Hamiltonian of the system. We define for $\hat{\cal{O}}$ a corresponding emergent QND observable as
\begin{align}
\hat{\mathcal{O}}_{\rm eQND} \equiv \sum_n \ket{n}\bra{n}\hat{\mathcal{O}}\ket{n}\bra{n},
\end{align}
with $\ket{n}$  energy eigenstates. Measurement of $\hat{\mathcal{O}}_{\rm eQND}$ provides the same information as of $\hat{\mathcal{O}}$ for the energy eigenstates, but in a non-destructive way. 

Following this definition, we immediately see that $\hat{f}_{z_0}^{(0)}$ is the eQND observable corresponding to $\hat{f}_{z_0}(\hat{z})$. In particular, in the limit $\sigma\to 0$, we have $f_{z_0}(\hat{z})\sim \ket{z_0}\bra{z_0}$ such that $\hat{f}_{z_0}^{(0)}\sim \sum_n|\braket{n|z_0}|^2\ket{n}\bra{n}$ directly provides the spatial density of energy eigenstates at the focal point $z_0$. Since $[\hat{f}_{z_0}^{(0)},\hat{H}_{A,E}]=0$, this measurement is non-destructive.

\begin{figure}[t!]
\centering{}\includegraphics[width=0.49\textwidth]{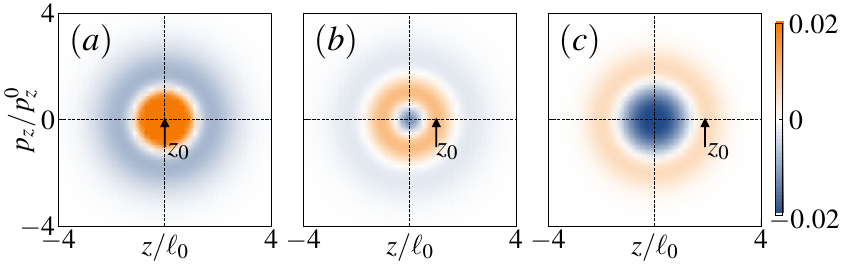} \caption{The measurement backaction in the {Scanning Mode} of the microscope. We choose $\sigma=0.5\ell_0$, and plot the Wigner function of $\mathcal{H}[f_{z_0}(\hat{z})]\tilde{\rho}_{\rm th}$ for a thermal state $\tilde{\rho}_{\rm th}=\sum_{n}p_{n}\ket{n}\bra{n}$ of a HO, with $p_n\propto e^{-n/n_{\rm th}}$ and $n_{\rm th}=1$, for focal point (a) $z_0=0$, (b) $z_0=\ell_0$ and (c) $z_0=2\ell_0$. The Wigner function shows a density flow in the phase space, corresponding to a redistribution of the population of the energy eigenstates (see text).}
\label{fig:figure8} 
\end{figure}

To gain an intuition of the backaction associated with this eQND measurement, in Fig.~\ref{fig:figure7} we plot the Wigner function of $\hat{f}_{z_0}^{(0)}$, which shows a symmetric distribution around the phase space center with a finite spread $\sim\sigma$. As such, continuous measurement of $\hat{f}_{z_0}^{(0)}$ does not lead to momentum diffusion. Rather, as in the familiar QND measurement~\cite{Gleyzes2007,Johnson2010,Volz2011,Barontini2015,Moller2017}, it reduces the coherence between different energy eigenstates, and selects out a particular energy eigenstate. This is illustrated in Fig.~\ref{fig:figure8}, where we show the effect of the homodyne term $\mathcal{H}[f_{z_0}(\hat{z})]$ acting on a thermal state of the atom in an HO potential, which induces redistribution of the population of motional eigenstates. 
%Consequently, depending on the Wiener increments $dW(t)$ [c.f., Eq.~\eqref{eq:SME_good_cavity}] in a particular measurement run, the measurement backaction %projects the atom onto  one of the energy eigenstates. 
Moreover, depending on the focal point $z_0$, the measurement backaction mainly changes the population of a particular eigenstate ${\ket{n}}$ with the largest matrix element $\bra{n}f_{z_0}(\hat{z})\ket{n}$. This can be clearly seen by comparing Fig.~\ref{fig:figure8}(a-c). 

The eQND measurement shares the same merit as the standard QND measurement~\cite{Clerk2010}: once the atomic state is projected onto an energy eigenstate, the only noise source in the measured homodyne current is the photon shot noise, which can be made arbitrarily small by increasing the measurement strength $\gamma$ (or equivalently, the measurement time). The extra decoherence terms in Eq.~\eqref{eq:SME_good_cavity} introduce slight imperfections to this ideal scenario. These include $\mathcal{D}[\hat{f}_{z_0}^{\ell}],\,\ell\neq0$, which describes incoherent quantum jumps between energy eigenstates, and the spontaneous decay term $\mathcal{L}_{\rm sp}$ qualitatively analyzed in Sec.~\ref{sec:Lsp}. The presence of these imperfections introduces extra noise to the homodyne current, thus reducing its SNR. Nevertheless, the features of the eQND measurement, in particular the ability to map out the spatial density of energy eigenstates with a high SNR in a single measurement run, is robust against these imperfections, as we show below.

\begin{figure*}[t!]
\centering{}\includegraphics[width=\textwidth]{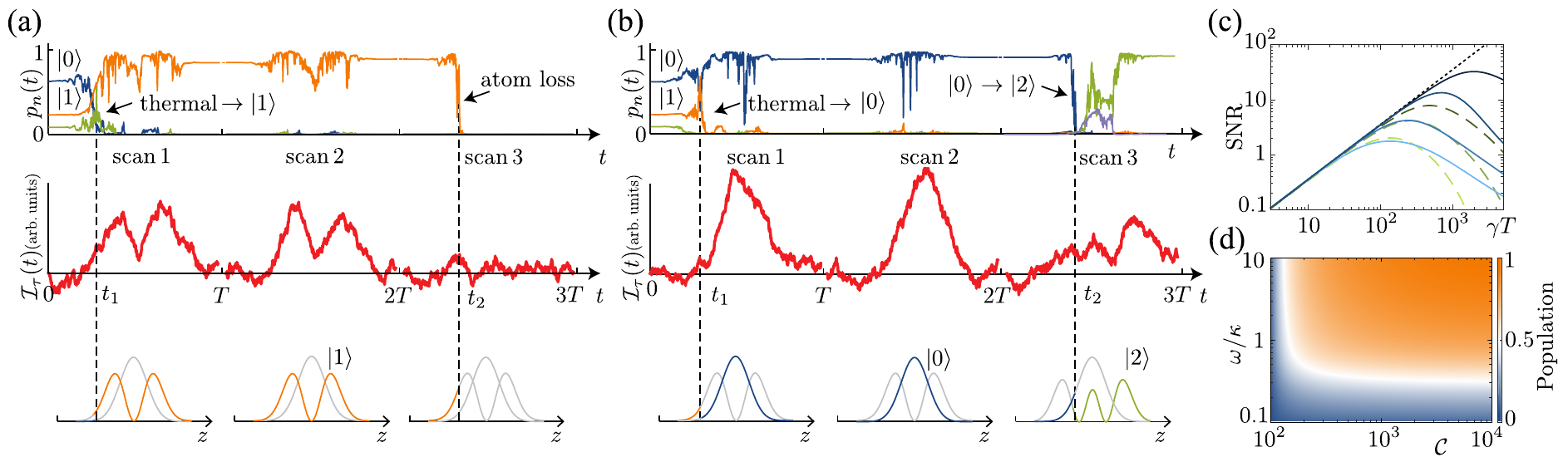} \caption{Single-run scans in the {Scanning Mode} of the microscope. (a) and (b) shows the simulation of two independent measurement runs for an atom initialized in a thermal state of a HO with $n_{\rm th}=1$ (see text). Each measurement run consists of three consecutive scans, where the focal point is moved across the atomic system in a time $T$. The conditional populations $p_n(t)$ ($n=0,1,2$) of the energy eigenstates (upper panel) and the filtered homodyne current ${\cal I}_{\tau}(t)$ (middle pannel) are shown together with the spatial density inferred from the trap populations (lower panel). The eQND measurement prepares the atom in a trap eigenstate [$\ket{1}$ in (a) and $\ket{0}$ in (b)] at a time labeled by $t_1$ in the first scan $t\in[0,T]$, and ${\cal I}_{\tau}(t)$ traces the corresponding density faithfully in the second scan $t\in[T,2T]$. In (a), the atom is lost at a time $t_2\in[2T,3T]$ due to spontaneous decay; whereas in (b), the atom jumps to the trap state $\ket{2}$ at $t_2\in[2T,3T]$. Parameters: $\gamma T = 1000$, $\mathcal{C}=200$, $\sigma=0.5\ell_0$. 
(c) SNR vs. $\gamma T$ for a single scan of an atom initialized in the eigenstate $\ket{1}$ of the HO, compared to an ideal QND measurement, Eq.~\eqref{eq:SNR_analytic} (dotted straight line), for $\sigma=0.5\ell_{0}$. SNR is
taken at $z_{0}(t)=-\ell_{0}$ (theoretical maximum). Solid curves: $\omega/\kappa= 10,4,1,0.1$ (dark to light), assuming no spontaneous emission (i.e., $\mathcal{C}=\infty$). Dashed curves: $\mathcal{C}=1000,200,100$ (dark to light), $\omega/\kappa=10$.
(d) For an atom initialized in $\ket{1}$, we plot its final population in $\ket{1}$ after completing a single scan, averaged over all measurement outcomes. We choose $\gamma T = 1000$ and $\sigma=0.5 \ell_0$. Increasing $\omega/\kappa$ and improving $\mathcal{C}$ greatly suppresses the population depletion.
}
\label{fig:figure9} 
\end{figure*}

\subsubsection{The stochastic rate equation}
To provide a physical interpretation for the dynamics of eQND measurement and the impact of imperfections,
let us expand Eq.~(\ref{eq:SME_good_cavity})
in the energy eigenbases and obtain a (nonlinear) stochastic
rate equation (SRE) for the eigenstate populations $\smash{p_{n}\equiv\langle n|\tilde{\rho}_{c}|n\rangle}$:
\begin{align}
\label{eq:SRE}
dp_{n}=&\sum_{\ell\neq0}A_{n}^{\ell} (p_{n+\ell} - p_{n})dt - B_{n} p_{n} \,dt \nonumber \\
&+ 2\sqrt{\gamma}p_{n}\left(f_{nn} - \sum\nolimits _{m}f_{mm}\,p_{m}\right)dW(t).
\end{align}
Here we have defined the rates 
\begin{align}
A_{n}^{\ell} &\equiv \left[\frac{\gamma}{1+(2\ell\omega/\kappa)^2} + \frac{\gamma}{4\mathcal{C}}\right]|f_{n,n+\ell}|^2,\label{eq:coef A}\\
B_{n} &\equiv \frac{\gamma}{4\mathcal{C}} \bra{n}f^2_{z_0}(\hat{z})\left[\frac{\Omega_r(\hat{z})}{\Omega_g}\right]^2\ket{n},\label{eq:coef B}
\end{align}
and have dropped small terms $\propto P_{re(ge)}$ as well as the fast-rotating terms in $\mathcal{L}_{\rm sp}$.

The effect of the measurement is captured by the stochastic term in the second line of Eq.~\eqref{eq:SRE}. It describes the collapse of the motional density matrix of the atom to a particular energy eigenstate,
$\tilde{\rho}_{c}(t)\rightarrow\ket{n}\bra{n}$, within a collapse
time $T_{\text{coll}}\sim1/\gamma$. 
The impact of higher sideband transitions and the atomic spontaneous emission is contained in the first line of Eq.~\eqref{eq:SRE}.  Here the first term describes redistribution of the population between the energy eigenstates, which preserves the total population $\sum_np_n$. In the limit of $\kappa/\omega\ll1$ and $\mathcal{C}\gg1$ this process happens on a much longer time scale, the dwell time $T_{\rm dwell}\sim [\gamma/4\mathcal{C}+\gamma(\kappa/2\omega)^2]^{-1}\gg T_{\rm coll}$. The second term describes the decay of the population of the motional eigenstates due to spontaneous emission. It also happens on a much longer time scale, $T_{\rm sp}\sim \gamma/4\mathcal{C}\gg T_{\rm coll}$.

As a result, the evolution of the atomic system corresponding to Eq.~\eqref{eq:SRE}
consists of a rapid collapse to an
energy eigenstate $\ket{n}$, followed by a sequence of rare quantum
jumps between the energy eigenstates on the time scale $T_{\text{dwell}}$, or loss of the atom on the time scale $T_{\text{sp}}$. Such a time scale separation allows us to define a time window $T_{\text{coll}}\ll T\lesssim T_{\text{dwell}}, T_{\rm sp}$, during which the information of the energy eigenstates can be extracted backaction free. This enables the Scanning Mode of the microscope, as discussed in the following.

\subsubsection{Application: preparing and scanning motional eigenstates}
\label{sec:simulation_good_cavity}
The Scanning Mode operates by moving
the focal point $z_{0}(t)$ across the atomic system, $-L/2<z_{0}(t)<L/2$, during a time $T$ satisfying $T_{\text{coll}}\ll T\lesssim T_{\text{dwell}}, T_{\rm sp}$. By starting the scan, the motional state of the atom will first collapse to a particular energy eigenstate $\ket{n}$ (note this stage can be viewed as state preparation), with the subsequent scan reading out the spatial density
profile $\langle n|\hat f_{z_{0}}^{(0)}|n\rangle=\int dz\,f_{z_{0}}(z)|\langle z|n\rangle|^{2}$, until the atom jumps to another energy eigenstates or gets lost due to spontaneous decay.

As an illustration, we consider the scan of an atom trapped in a HO potential. To this end, we choose $\omega/\kappa=10$ and assume that the atom is prepared at $t=0$  in a thermal motional state of the HO potential, $\tilde{\rho}(0)=\tilde{\rho}_{\rm th}=\sum_{n}p_{n}\ket{n}\bra{n}$ with $p_n\propto e^{-n/n_{\rm th}}$ and $n_{\rm th}=1$. In our illustration, we perform three consecutive spatial scans covering
$-L/2<z_{0}(t)<L/2$ ($L=8\ell_{0}$), each for a time interval $T$ ($\gamma T=1000$). In Fig.~\ref{fig:figure9}(a), we show the resulting homodyne current and the corresponding population of the energy eigenstates \emph{in a single run}, based on integrating the SRE~(\ref{eq:SRE}). In filtering the homodyne current, we choose the `optimal' filter time $\tau = \sigma / v$, with $\sigma$ the microscope resolution and $v=L/T$ the scanning speed. As can be seen in Fig.~\ref{fig:figure9}(a), in \emph{scan 1} the microscope prepares the atom in a particular energy eigenstate (here the first excited state $\ket{1}$) in a random way according to the initial state distribution. The atom stays in $\ket{1}$ in \textit{scan 2}, allowing for a faithful readout of its spatial density through the detected homodyne current. In \textit{scan 3}, the atom stays in $\ket{1}$ until it suddenly gets lost due to spontaneous decay out of its internal dark state, manifesting in the homodyne current as a sudden jump to zero. Such a loss event is fast (on a time scale $\sim T_{\rm coll}$) but rare (on a time scale $\sim T_{\rm sp}$ after the beginning of scan). In Fig.~\ref{fig:figure9}(b), we show another simulation 
of Eq.~\eqref{eq:SRE} 
representing another independent measurement run. In this run, the microscope prepares the atom in the motional ground state $\ket{0}$ in \emph{scan 1}, and subsequently reveals its spatial density in \emph{scan 2}. 
But in \emph{scan 3} of Fig.~\ref{fig:figure9}(b), the atom first stays in $\ket{0}$, then instead of disappearing suddenly jumps to the second excited state $\ket{2}$, with the homodyne current starting tracing the density profile of $\ket{2}$. Such a quantum jump is induced by the higher sideband transition terms $\mathcal{D}[\hat{f}_{z_0}^{\ell}],\,\ell\neq0$. It is fast (on a time scale $\sim T_{\rm coll}$) but rare ($\sim T_{\rm dwell}$ between adjacent jumps).

To quantify the performance of the scanning microscope in the presence of imperfections, 
 we plot in Fig.~\ref{fig:figure9}(c) the SNR of a single scan of a motional eigenstate of the atom as a function of the (dimensionless) measurement strength $\gamma T$ for different $\omega/\kappa$ and different cooperativity $\mathcal{C}$. The solid curves reflects the effect of higher sideband transitions, $\mathcal{D}[\hat{f}_{z_0}^{\ell}]$ in Eq.~\eqref{eq:SME_good_cavity}, with $\mathcal{L}_{\rm sp}=0$. As can be seen, for small $\gamma$ the SNR increases with $\gamma$ linearly (i.e., in a QND fashion). For large $\gamma$, however, the SNR deviates from linear increase due to these higher sidebands transitions. By increasing $\omega/\kappa$ we greatly suppress these processes, rendering them into rarer quantum jumps, thus improving the SNR. The effect of atomic spontaneous decay is shown as the dashed curves in Fig.~\ref{fig:figure9}(c), where we plot the behavior of the SNR for different $\mathcal{C}$ while keeping $\omega/\kappa$ fixed. As expected, spontaneous decay degrades the achievable SNR, as it brings the atomic population out of the dark state. This effect is, however, strongly suppressed for large $\mathcal{C}$.

As another indicator of the microscope performance, we plot in Fig.~\ref{fig:figure9}(d)  the remaining population of an initially populated motional eigenstate after completing a single scan, averaged over all measurement runs. As can be seen, in the regime $\kappa\ll\omega$ and $\mathcal{C}\gg 1$, the population remains around $1$, indicating a nearly ideal eQND measurement.

To summarize, Fig.~\ref{fig:figure9} demonstrates that by taking a good cavity $\kappa\ll\omega$ and choosing sufficiently large cooperativity $\mathcal{C}\gg 1$ to suppress the atomic spontaneous emission, the scanning mode of the microscope is able to map out the spatial density of energy eigenstates with a high SNR in a single scan as an eQND measurement.

\subsubsection{The resolution limit of the microscope}
\label{sec:resolution_limit}

As it was already mentioned in the Sec.~\ref{sec:focusingFunction},
the spatial resolution of the microscope is limited by the spontaneous
decay processes leading to the loss of an atom. In this section we
estimate analytically and evaluate numerically the effect of the spontaneous
emission on the SNR in the Scanning Mode, as a function of the resolution
$\sigma$.

In the limit of high spatial resolution $\sigma\ll\ell_{0}$ with
$\ell_{0}$ being the characteristic length of the atomic wavefunction,
the focusing function $f_{z_{0}}(z)$ has the form of a narrow peak
with the height $\sim\ell_{0}/\sigma$ and the width $\sim\sigma$
around $z_{0}$. After assuming that the system during the scan has
already collapsed to an eigenstate $\ket{n}$ and averaging the photocurrent
in Eq.~\eqref{eq:homodyne_good_cavity_maintext} over the time window
$\tau=T\sigma/L$, we obtain an estimate for the SNR limited by the shot
noise: 
\begin{equation}
\mathrm{SNR}(z_{0},T)\simeq4\gamma T\frac{\sigma}{L}\big|\tilde{\psi}_{n}({z_{0}}/{\ell_{0}})\big|^{4},\label{eq:SNR_analytic}
\end{equation}
where we define the dimensionless wavefunction $\tilde{\psi}_{n}({z_{0}}/{\ell_{0}})=\sqrt{\ell_{0}}\braket{n|z_{0}}$.
The linear dependence of the SNR on the total scan
time $T$ is shown as a dotted line in Fig.~\ref{fig:figure9}(c).
However, for long enough $T$, the decoherence processes become important
and result in the deviation from the linear dependence. In the limit
$\sigma\ll\ell_{0}$, the corresponding time scale is defined by the
$B_{n}$ terms in Eq.~\eqref{eq:SRE}, which, following Eq. (\ref{eq:coef B}),
can be estimated as 
\[
B_{n}(z_{0})\sim\frac{\gamma}{\mathcal{C}}\frac{E_{r}}{V_{{\rm na}}^{\mathrm{max}}}\frac{\ell_{0}}{k_{0}^{2}\sigma^{3}}\big|\tilde{\psi}_{n}({z_{0}}/{\ell_{0}})\big|^{2}.
\]
The terms $A_{n}^{\ell}$ can be neglected because the matrix elements
$|f_{nm}|$ are of order 1 for any $\sigma$ and, therefore, $A_{n}^{\ell}\ll B_{n}$
for $\sigma\ll\ell_{0}$ and $\omega/\kappa\gg1$. As a result, for the total scan
over the distance $L$ during the time $T$, the atom loss probability
reads
\begin{equation}
\widetilde{B}_{n}T\sim\frac{\gamma T}{\mathcal{C}}\frac{E_{r}}{V_{{\rm na}}^{\mathrm{max}}}\frac{1}{k_{0}^{2}\sigma^{3}}\frac{l_{0}^{2}}{L},\label{eq:measurement_time_restriction}
\end{equation}
where $\widetilde{B}_{n}=L^{-1}\int B_{n}(z_{0})dz_{0}$ is the spatial
average. For $\widetilde{B}_{n}T\ll$1, the atom loss can be neglected
and the SNR grows linearly with $T$ following Eq.~\eqref{eq:SNR_analytic}.
For longer $T$, however, the atom can eventually be lost with the
result that the average photocurrent drops to zero which we regard
as a noise. Therefore, after taking the effect of the atomic spontaneous
decay into account, the SNR can be written as 
\begin{equation}
\mathrm{SNR}(z_{0},T)\simeq\frac{4\gamma T\frac{\sigma}{L}\big|\tilde{\psi}_{n}({z_{0}}/{\ell_{0}})\big|^{4}}{1+4\gamma T\frac{\sigma}{L}\big|\tilde{\psi}_{n}({z_{0}}/{\ell_{0}})\big|^{4}\widetilde{B}_{n}T},\label{eq:SNR_analytic_full}
\end{equation}
which is in a good agreement with the dashed lines in Fig.~\ref{fig:figure9}(c)
that represent the direct numerical simulations of Eq.~\eqref{eq:SRE}.

\begin{figure}[t!]
\centering{}\includegraphics[width=0.49\textwidth]{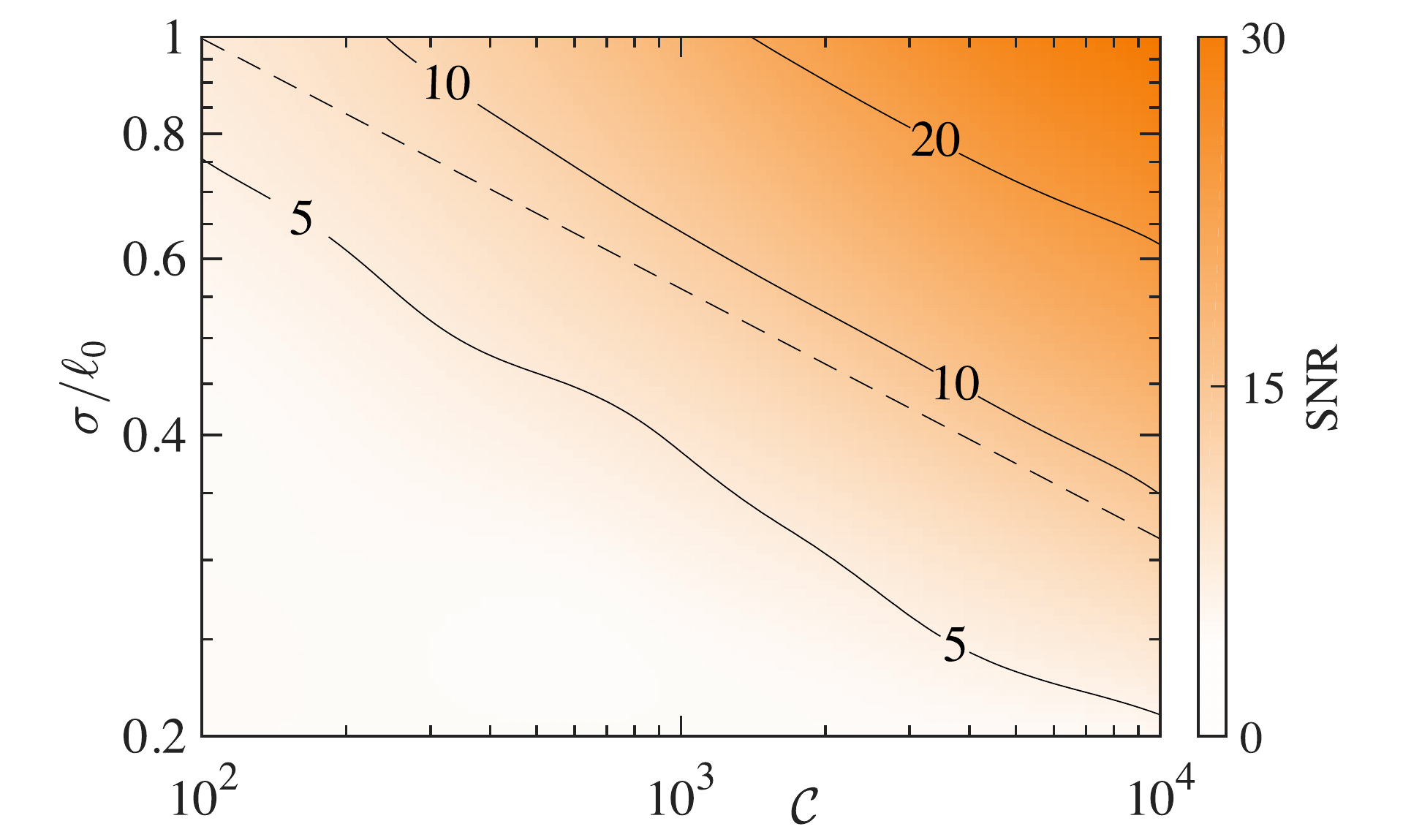} \caption{The resolution limit of the microscope. The relation between the cavity
cooperativity $\mathcal{C}$, the spatial resolution $\sigma$ (in
units of the HO length $\ell_{0}\equiv\sqrt{\hbar/m\omega}$), and
the maximally achievable SNR for a single scan of an atom initialized
in $\ket{1}$ of a HO potential (SNR is calculated for the maximum
of the wave function at $z_{0}(t)=-\ell_{0}$ and optimized over $\gamma T$)
. Parameters are $\omega/\kappa=10$ and $V_{\mathrm{na}}^{\mathrm{max}}=0.1\hbar\omega$.
Dashed line shows the scaling $\sigma/\ell_{0}\sim\mathcal{C}^{-1/4}$
according to Eq.~\eqref{eq:resolution_scaling} for a fixed SNR.}
\label{fig:figure10} 
\end{figure}

By evaluating the SNR at the maximum of the wavefunction and optimizing
it over the measurement time $T$ in Eq.~\eqref{eq:SNR_analytic_full},
we obtain a universal expression 
\begin{equation}
\left(\frac{\sigma}{\lambda_{0}}\right)_{\mathrm{min}}\sim\left(\frac{\mathrm{SNR}^{2}}{\mathcal{C}}\frac{E_{r}}{V_{{\rm na}}^{\mathrm{max}}}\frac{\ell_{0}^{2}}{\lambda_{0}^{2}}\right)^{1/4}\label{eq:resolution_scaling}
\end{equation}
for the resolution limit of the microscope. The corresponding optimal
measurement time is given by $\gamma T\sim k_{0}\sigma(L/\ell_{0})\sqrt{\mathcal{C}\,V_{{\rm na}}^{\mathrm{max}}/E_{r}}$.
It follows from Eq.~\eqref{eq:resolution_scaling} that the spatial
resolution is limited by the cavity cooperativity $\mathcal{C}$ and
by the chosen SNR (of a single measurement run). Improving the resolution
leads to reducing the SNR, which, however, can be compensated by
increasing the cavity cooperativity $\mathcal{C}$ to suppress the
atomic spontaneous emission. Figure~\ref{fig:figure10} shows the
relation between the maximally achievable SNR, the spatial resolution,
and the cavity cooperativity, calculated from numerical simulation
of Eq.~\eqref{eq:SRE} for the case of the harmonic oscillator, which
is in a good agreement with the scaling behavior predicted by Eq.~\eqref{eq:resolution_scaling}.

\section{Experimental Feasibility}
\label{sec:exp_considerations} 
We now address the experimental feasibility of the quantum scanning microscope.
As we will show below, the present state of art experiments provide
all necessary ingredient for realization of the microscope itself
and both operation modes.

First of all, the proposed setup for quantum scanning microscope requires
trapping of atoms inside a high-$Q$ optical
cavity, which has already been realized in several experimental platforms
ranging from cold neutral atoms in optical traps (lattices)~\cite{Ritsch2013}
to trapped ions~\cite{Northup2014}. Another ingredient, the homodyne
detection, is a well-developed technique that can be performed with
nearly unit efficiency~\cite{Krauter2013}. An implementation of
the subwavelength focusing function via the dark state engineering
in an atomic $\Lambda$ configuration, cf. Fig.~\ref{fig:figure1},
also looks realistic in view of the recent experimental realization~\cite{PhysRevLett.120.083601}
of the subwavelength optical barriers and the possibility to shine
additional lasers from the side as was done in~\cite{PhysRevLett.114.023602}.

The Movie Mode of the microscope, Sec.~\ref{badCavityLimit}, requires
the bad cavity condition $\kappa\gg\omega$. In fact, this is the typical situation for cavity
QED experiments with optically trapped neutral atoms (see, e.g, Ref.~\cite{Ritsch2013}),
where the cavity linewidth $\kappa\sim$ MHz is much larger
than the frequency of the atomic motion $\omega\sim E_{r}/\hbar\sim$
kHz.

The Scanning Mode, Sec.~\ref{sec3}, needs a good cavity with $\kappa\ll\omega$.
This condition can be met in cavity QED setups with optically trapped
neutral atoms. For example, Ref.~\cite{Wolke75} reports strong coupling
between an atomic BEC with a narrow line width optical cavity with
$\kappa\simeq2\pi\times4.5{\rm kHz}$, far smaller than the optical
recoil energy for light-mass alkalies (e.g., $E_{r}/\hbar\simeq2\pi\times23{\rm {kHz}}$
for $^{23}$Na at the D$2$ line with $\lambda_{0}\simeq590$nm).
Ions trapped in optical cavities \cite{PhysRevLett.114.023602} provide another
feasible platform for reaching the good cavity condition due to their
large oscillation frequency ($\omega\sim$~MHz).

The spontaneous emission, as discussed in Sec.~\ref{sec:Lsp}, degrades
the measured homodyne current due to gradual depletion of atom from
the dark state. This detrimental effect can be strongly suppressed
by using high-$Q$ optical cavities with large cooperativity (e.g.,
$\mathcal{C}>100$ in Ref.~\cite{Kimble1998}).

For a concrete illustration, we consider the example
of a single $^{23}$Na atom trapped in an optical lattice with the
amplitude $V_{0}=5E_{r}$, where $E_{r}/\hbar\simeq2\pi\times23{\rm {kHz}}$,
which corresponds to the harmonic oscillator frequency $\omega=\sqrt{2V_{0}E_{r}}/\hbar\simeq2\pi\times76$kHz
and the size of the ground state $\ell_{0}=\sqrt{\hbar/m\omega}\simeq75$nm.
For the focusing function we choose $\epsilon=\beta/3 = 0.1$, which
leads to a resolution {[}see Eq.~\eqref{appResolution}{]} $\sigma\simeq 0.07\lambda_{0}\simeq0.5\ell_{0}$
and $V_{\rm na}^{\rm max}\simeq0.1\hbar\omega$ {[}see Eq.~\eqref{nonAdiabaticCorrection}{]}
being much smaller than the level spacing in the trap. For a cavity
with the cooperativity $\mathcal{C}=200$ we have for the spontaneous
emission rate $\sim\gamma/4\mathcal{C}\ll\gamma$, which is negligible
for the Movie mode, cf. Fig.~\ref{fig:figure6}, and provides high
enough SNR, cf. Fig.~\ref{fig:figure9}, to map out the atomic density
distribution in a single experimental run in the Scanning Mode.

\section{Conclusion and outlook}
\label{sec:conclusion}
In conclusion, we have presented a detailed theoretical description of the quantum scanning microscope for cold atoms in the CQED setup proposed in \cite{Yang2017a}, and discussed its experimental feasibility. The microscope is conceptually different from the familiar destructive `microscopes' in cold atom experiments by allowing a continuous monitoring of an atomic system with optical subwavelength spatial resolution and by demonstrating the nondemolition observation of the atomic density operator as an emergent QND measurement. The concept of the emergent QND measurement extends the notion of the backaction free continuous measurement to the case of a general quantum mechanical observable monitored in a system energy eigenstate.
 
 Furthermore, we have demonstrated the action of the microscope as a device for continuous observation with two examples illustrating the two different operation modes: the observation of the atomic wave packet moving in a harmonic trap through the fixed focal region (the Movie Mode), and a scan of the atomic density distribution in a motional eigenstate of a harmonically trapped atom by moving the focal region slowly across the trap (the Scanning Mode). In the latter case, the microscope can be used for a probabilistic preparation of the atomic system in a pure motional eigenstate starting from an initial mixed one as a result of the measurement induced state collapse. These examples demonstrate the fundamental difference in the action of the proposed microscope allowing continuous observation of a quantum system from the common measurement scenario with the quantum gas microscope where a single shot destructive measurement terminates a given run of the experiment.
 
We also mention that the ideas behind the microscope operation and the emergent QND measurement are not necessarily restricted to  CQEDs implementation considered here, but can be realized with other experimental platforms, e.g. coupling the atom of interest to an ensemble of Rydberg atoms for read out. Finally, we emphasize that continuous observation of a quantum system provides the basis of a quantum feedback~\cite{wiseman2009quantum} on the system of interest, which is of particular interest for quantum many-body systems.

\subsection*{Acknowledgements}
We acknowledge J. Reichel for fruitful discussions and comments. Part of the numerical simulations were performed using the QuTiP library~\cite{JOHANSSON20131234}. Work at Innsbruck is supported
by the Austrian Science Fund SFB FoQuS (FWF Project No. F4016-N23) and 
the European Research Council (ERC) Synergy Grant UQUAM. 

\bibliography{scanningMicroscope_bib}
\newpage

\begin{appendix}
\section{Details on the adiabatic elimination of the atomic internal dynamics}
\label{app2}
In this appendix we present details on the adiabatic elimination of the atomic internal DOFs, which is sketched briefly in Sec.~\ref{sec:elimination_internal} of the main text. As described in Sec.~\ref{sec:elimination_internal}, we are interested in the regime where (i) the external motion of an atom is much slower than its internal dynamics, $|\hat{H}_{A,E}| \ll |\hat{H}_{A,I}| $ and (ii) the atom is coupled to the cavity mode dispersively, $g(z)\ll\Delta_t$. Condition (i) allows us to define a small parameter $\varepsilon_{1}={\rm Tr}(\mu_c\hat{H}_{A,E})/\hbar\Omega(z)$, where $\mu_c$ is the density matrix for the joint atom-cavity system introduced in Sec.~\ref{sec:SME_full}, and $\Omega(z)\equiv\sqrt{\Omega_g^2+\Omega_r^2(z)}$ characterizes the energy gap between the internal dark and bright states. Condition (ii) allows us to define another small parameter $\varepsilon_{2}={\rm Tr}(\mu_c\hat{H}_{AC})/(\hbar\Delta_t)$. Our aim below is to derive an equation for the reduced density matrix $\rho_c\equiv {\rm Tr}_{A,I} (\mu_c)$ describing the cavity mode and the atomic external motion, where ${\rm Tr}_{A,I}$ is a trace over the atomic internal DOFs, by perturbation theory to Eq.~\eqref{completeSMEmaintext} in terms of the small parameters $\varepsilon_{1,2}$. The desired equation has deterministic terms accurate to second order in the perturbation $\varepsilon_{1,2}$, and a stochastic term accurate to linear order in the perturbation $\varepsilon_{1,2}$, i.e.,
\begin{equation}
\label{perturbativeExpression}
d\rho_{c}=O(\varepsilon_1^2,\varepsilon_2^2)dt+O(\varepsilon_1,\varepsilon_2)dW(t).
\end{equation}
To this end, in the following we first analyze the structure of Eq.~\eqref{completeSMEmaintext} in the internal dark and bright state basis, then carry out the adiabatic elimination.

\subsection{Stochastic Master Equation in the Dark and Bright State Basis}
In order to perform the adiabatic elimination, we express the SME~\eqref{completeSMEmaintext} of the main text in terms of $\{\ket{D},\ket{\pm},\ket{t}\}$, i.e., the eigenstates of the atomic internal Hamiltonian $\hat{H}_{A,I}$, defined in Eqs.~\eqref{eq: dark_state_def} and~\eqref{eq: bright_state_def} of the main text. In terms of them, we have
\begin{align}
\label{eq: internal_states_app}
\ket{e}&= \frac{1}{\sqrt{2}}\left(\ket{+} -\ket{-}\right),\nonumber\\
\ket{g} &=\frac{1}{\sqrt{2}}\cos\alpha(z)(\ket{+}+\ket{-})+\sin\alpha(z) \ket{D},\nonumber\\
\ket{r} &=\frac{1}{\sqrt{2}}\sin\alpha(z)(\ket{+}+\ket{-})-\cos\alpha(z) \ket{D},\nonumber\\
\end{align}
with the mixing angle $\alpha(z)$ defined in Eq.~\eqref{eq:mixing angle 1} of the main text. Eq.~\eqref{eq: internal_states_app} allows us to express each term of the SME~\eqref{completeSMEmaintext} in the new basis.

The atomic spontaneous emission terms (i.e. the last two lines) of Eq.~\eqref{completeSMEmaintext} can be readily simplified in this new basis, under the condition $\Gamma_t,\Gamma_e \ll\Omega(z),|\Delta_t|,|\Omega(z)/2\pm\Delta_t|$. This condition allows us to neglect the fast rotating terms in the spontaneous emission channels under the rotating-wave approximation. As a result, Eq.~\eqref{completeSMEmaintext} can be expressed as
%\begin{align}
%\label{completeSMEsecular}
%d\mu_c =&- \frac{i}{\hbar}[\hat{H}_{A,I}+\hat{H}_{A,E}+\hat{H}_C+\hat{H}_{AC},\mu_c ] dt\nonumber\\
%&+ \kappa\mathcal{D}[\hat{c}]\mu_c dt + \sqrt{\kappa}\mathcal{H}[\hat{c}e^{-i\phi}]\mu_c dW(t)\nonumber\\
%&+\Gamma\int du N(u) \mathcal{D}[e^{-i k u \hat{z}}\cos\alpha(\hat{z})\hat{\sigma}_{Dt}]\mu_cdt\nonumber\\
%&+\frac{\Gamma}{2}\sum_{j=\pm}\int du N(u) \mathcal{D}[e^{-i k u \hat{z}}\sin\alpha(\hat{z})\hat{\sigma}_{jt}]\mu_cdt\nonumber\\
%&+\frac{\Gamma P_g}{2}\sum_{j=\pm}\int du N(u) \mathcal{D}[e^{-i k u \hat{z}}\sin\alpha(\hat{z})\hat{\sigma}_{Dj}]\mu_cdt \nonumber\\
%&+\frac{\Gamma P_r}{2}\sum_{j=\pm}\int du N(u) \mathcal{D}[e^{-i k u \hat{z}}\cos\alpha(\hat{z})\hat{\sigma}_{Dj}]\mu_cdt\nonumber\\
%&+\frac{\Gamma P_g}{4}\sum_{j,l=\pm}\int du N(u) \mathcal{D}[e^{-i k u \hat{z}}\cos\alpha(\hat{z})\hat{\sigma}_{jl}]\mu_cdt\nonumber\\
%&+\frac{\Gamma P_r}{4}\sum_{j,l=\pm}\int du N(u) \mathcal{D}[e^{-i k u \hat{z}}\sin\alpha(\hat{z})\hat{\sigma}_{jl}]\mu_cdt\nonumber\\
%&- \frac{\Gamma P_o}{4}\sum_{j=\pm} \{ \hat{\sigma}_{jj},\mu_c\}dt,
%\end{align}
\begin{align}
\label{completeSMEsecular}
d\mu_c =&- \frac{i}{\hbar}[\hat{H}_{A,I}+\hat{H}_{A,E}+\hat{H}_C+\hat{H}_{AC},\mu_c ] dt\nonumber\\
&+ \kappa\mathcal{D}[\hat{c}]\mu_c dt + \sqrt{\kappa}\mathcal{H}[\hat{c}e^{-i\phi}]\mu_c dW(t)\nonumber\\
&+\frac{\Gamma_t}{2}\sum_{j=\pm}\mathcal{G}_2^{rt}[\hat{\sigma}_{jt}]\mu_c dt\nonumber\\
&+\frac{\Gamma_e}{2}\sum_{j=\pm}\left(P_{ge}\mathcal{G}_2^{ge}[\hat{\sigma}_{Dj}]\!+\!P_{re}\mathcal{G}_1^{re}[\hat{\sigma}_{Dj}]\right)\!\mu_cdt\nonumber\\
&+\frac{\Gamma_e}{4}\sum_{j,l=\pm}\left(P_{ge}\mathcal{G}_1^{ge}[\hat{\sigma}_{jl}]+P_{re}\mathcal{G}_2^{re}[\hat{\sigma}_{jl}]\right)\mu_cdt\nonumber\\
&- \frac{\Gamma_e P_a}{4}\sum_{j=\pm} \{ \hat{\sigma}_{jj},\mu_c\}dt+\Gamma_t\mathcal{G}_1^{rt}[\hat{\sigma}_{Dt}]\mu_c dt.
\end{align}
Here to simplify the notation we have defined
\begin{align}
\label{eq:supG_appendix}
\mathcal{G}_1^{jn}[\bullet]\equiv\int du N_{jn}(u) \mathcal{D}[e^{-i k u \hat{z}}\cos\alpha(\hat{z})\,\bullet\,],\nonumber\\
\mathcal{G}_2^{jn}[\bullet]\equiv\int du N_{jn}(u) \mathcal{D}[e^{-i k u \hat{z}}\sin\alpha(\hat{z})\,\bullet\,],
\end{align}
with $j=r,g$, $n=e,t$ and $\bullet$ represending a general operator. In Eq.~\eqref{completeSMEsecular}, the atomic spontaneous emissions are described by the last three lines, which occur only between the internal (dressed) eigenstates as a result of the rotating-wave approximation.

Next, we consider the expression of the rest of the terms (i.e. the first two lines) of Eq.~\eqref{completeSMEsecular} in the new basis. We note that these terms are diagonal in $\{D,\pm,t\}$ except $\hat{H}_{A,E}$ and $\hat{H}_{AC}$, which we now analyze. We first look at $\hat{H}_{A,E}=\hat{p}_z^2/2m+V(\hat{z})$. Due to the position-dependence of the dark and bright states, the momentum operator $\hat{p}_z$ is non-diagonal in this basis and acquires an additional `gauge potential'. We can write it as 
\begin{equation*}
\hat{p}_{z}=\sum_{i,j\in\{D,\pm,t\}}\langle i|\hat{p}_{z}|j\rangle\hat{\sigma}_{ij}=-i\hbar\partial_{z}\otimes\mathbb{I}-\hat{A},
\end{equation*}
where the spatial derivative $\partial_z$ acts only on the motional DOFs of the atom, and $\mathbb{I}=\sum_{i\in\{D,\pm,t\}}\ket{i}\langle i|$ is the identity operator for the internal states,
\begin{equation*}
\hat{A}=\left[\frac{i}{\sqrt{2}}\alpha'(\hat{\sigma}_{+D}+\hat{\sigma}_{-D})+\rm{H.c.}\right]
\end{equation*}
is a position-dependent `vector potential' which couples the internal dark state to the bright states.
As a result, $\hat{H}_{A,E}$ is non-diagonal in the $\{D,\pm,t\}$ basis. Through straightforward calculation we find it can be written as the sum of two parts, $\hat{H}_{A,E}=\hat{H}_{A,E}^{\rm 0}+\hat{H}_{A,E}^{\rm 1}$. The first part doesn't couple the dark state to the bright states,
\begin{equation}
\label{eq:H_ext_app_diag}
\hat{H}_{A,E}^{\rm 0}=\left[-\frac{\hbar^2\partial^2_z}{2m}\!+\!V(\hat{z})\right]\otimes\,\mathbb{I}\,+V_{\rm na}(\hat{z})\,\otimes\bigg(\hat{\sigma}_{DD}+\frac{1}{2}\sum_{i,j=\pm}\!\hat{\sigma}_{ij}\bigg)
\end{equation}
Here we have defined an effective potential for the atomic external motion,
\begin{equation}
V_{\rm na}(\hat{z})\equiv\frac{\hbar^2}{2m}[\alpha'(\hat{z})]^2,
\end{equation}
which corresponds to the lowest order non-adiabatic correction to the atomic external motion due to the spatially varying internal states. The second term $\hat{H}_{A,E}^1$ couples the dark state to the bright states,
\begin{equation}
\begin{split}
\label{eq:H_ext_app_nondiag}
\hat{H}_{A,E}^{\rm 1}=&\frac{i\hbar}{2m}(\hat{A}\partial_z+\partial_z \hat{A})\\
=&\hat{H}_1\otimes(\hat{\sigma}_{+D}+\hat{\sigma}_{-D}+\rm{H.c.}),
\end{split}
\end{equation}
with
\begin{align}
\hat{H}_1=&-\frac{\hbar^2}{2\sqrt{2}m}(2\alpha'\partial_z+\alpha'').
\end{align}

Similarly, the atom-cavity coupling Hamiltonian $\hat{H}_{AC}=\hbar g(\hat{z})(\hat{c}^\dag \hat{\sigma}_{rt}+\mathrm{H.c.})$ can be written as $\hat{H}_{AC}=\hat{H}_{AC}^0+\hat{H}_{AC}^1$, here
\begin{equation}
\hat{H}_{AC}^0=\frac{1}{\sqrt{2}}\hbar g(\hat{z})\sin\alpha(\hat{z})(\hat{c}\hat{\sigma}_{-t}+\hat{c}\hat{\sigma}_{+t}+{\rm H.c.})
\end{equation}
doesn't couple the dark state to the other states, while
\begin{align}
\label{eq: gcoupling_app}
\hat{H}_{AC}^1=\hbar\tilde{g}(\hat{z})\hat{c}\hat{\sigma}_{tD}+{\rm H.c.}
\end{align}
couples $\ket{D}$ to $\ket{t}$, with a strength 
\begin{equation}
\tilde{g}(\hat{z})\equiv-g(\hat{z})\cos\alpha(\hat{z}).
\end{equation}

In the following we shall eliminate the atomic internal dynamics by perturbation theory in terms of the coupling Hamiltonians Eq.~\eqref{eq:H_ext_app_nondiag} and \eqref{eq: gcoupling_app}.

\subsection{Adiabatic Elimination}
We now derive an effective equation of motion for the reduced density matrix $\rho_c\equiv {\rm Tr}_{A,I} (\mu_c)$. To this end, we trace out the atomic internal DOFs on both sides of Eq.~\eqref{completeSMEsecular}, yielding
\begin{align}
\label{darkstate_SME_1}
d\rho_{c} &= \mathcal{L}_0 \rho_{c}dt  +  \sqrt{\kappa}\mathcal{H}[\hat{c}e^{-i\phi}]\rho_c dW(t)\nonumber\\
&-\frac{i}{\hbar}[\hat{H}_1,\eta_++\eta_+^\dag+\eta_-+\eta_-^\dag]dt\nonumber\\
&-i\left([\,\tilde{g}(\hat{z})\hat{c}^\dag,\hat{\eta}_t]+[\,\tilde{g}(\hat{z})\hat{c},\hat{\eta}_t^\dag]\right)dt\nonumber\\
&+\frac{i}{\hbar}[V_{\rm na}(\hat{z}),\hat{\zeta}_t+\frac{1}{2}(\hat{\zeta}_++\hat{\zeta}_-)]dt+\Gamma_t (\mathcal{K}_1^{rt}+\mathcal{K}_2^{rt}-1)\hat{\zeta}_{t}dt\nonumber\\
&+\frac{\Gamma_e}{2} \Big[\sum_{j=g,r}P_{je}(\mathcal{K}_1^{je}+\mathcal{K}_2^{je})-1\Big](\hat{\zeta}_{+}+\hat{\zeta}_{-} )dt,
\end{align}
Here we have defined $\hat{\eta}_{i}\equiv{\rm Tr}_{A,I}(\hat{\sigma}_{Di}\mu_c)$ and $\hat{\zeta}_i\equiv{\rm Tr}_{A,I}(\hat{\sigma}_{ii}\mu_c)$, for $i\in\{t,\pm\}$, and have neglected terms $\propto{\rm Tr}_{A,I}(\hat{\sigma}_{\pm t}\mu_c)$ and $\propto{\rm Tr}_{A,I}(\hat{\sigma}_{+-}\mu_c)$. In Eq.~\eqref{darkstate_SME_1} we have defined a superoperator
\begin{align}
\label{eq:L0_app}
\mathcal{L}_0\rho_c=&-\frac{i}{\hbar}\left[-\frac{\hbar^2\partial^2_z}{2m}\!+\!V(\hat{z})\!+\!V_{\rm na}(\hat{z})\!+\!\hat{H}_{C},\rho_c\right]+ \kappa\mathcal{D}[\hat{c}]\rho_c
\end{align}
which includes all the Hamiltonians in Eq.~\eqref{completeSMEsecular} that doesn't couple the dark state to other internal states. 
We have also introduced the superoperators 
\begin{align}
\label{eq:supK_appendix}
\mathcal{K}_1^{jn}\bullet\equiv&\cos\alpha(\hat{z})\mathcal{K}^{jn}\bullet\cos\alpha(\hat{z})\nonumber\\
\mathcal{K}_2^{jn}\bullet\equiv&\sin\alpha(\hat{z})\mathcal{K}^{jn}\bullet\sin\alpha(\hat{z}),
\end{align}
where $\bullet$ stands for a general operator, and the superoperator $\mathcal{K}^{jn}\bullet=\int du N_{jn}(u) e^{-ik_0 u \hat{z}}\bullet e^{ik_0 u \hat{z}}$ describes the momentum diffusion for the spontaneous emission channel $\ket{n}\to\ket{j}$. 

To solve for $\rho_c$ in Eq.~\eqref{darkstate_SME_1}, we should determine $\hat{\eta}_i$ and $\hat{\zeta}_i$. It's easy to see that $\hat{\eta}_i\sim O(\varepsilon_1,\varepsilon_2)$ and $\hat{\zeta}_i\sim O(\varepsilon_1^2,\varepsilon_2^2)$. Thus, to achieve the accuracy prescribed by Eq.~\eqref{perturbativeExpression}, we need to know the evolution of $\hat{\eta}_i$ accurate to a linear order deterministic term in $\varepsilon_{1,2}$ and to a constant stochastic term $d\hat{\eta}_{i}=O(\varepsilon_1,\varepsilon_2)dt+O(1)dW(t)$. They can be derived straightforwardly from Eq.~\eqref{completeSMEsecular} as
\begin{align}
\label{eq:appendix_approximate_eta}
d\hat{\eta}_\pm=& \frac{1}{2}\left[\mp i\Omega(\hat{z})-\frac{\Gamma_e}{2}\right]\hat{\eta}_{\pm}dt+\mathcal{L}_0 \hat{\eta}_{\pm}dt - \frac{i}{\hbar}\hat{H}_1\rho_cdt,\nonumber\\
d\hat{\eta}_t=&\left(i\Delta_t-\frac{\Gamma_t}{2}\right)\hat{\eta}_tdt+\mathcal{L}_0 \hat{\eta}_{t}dt - i\tilde{g}(\hat{z})\hat{c}\rho_cdt,
\end{align}
where on the RHS we have used the fact ${\rm Tr}_{A,I}(\hat{\sigma}_{DD}\mu_c)=\rho_c+O(\varepsilon_1^2,\varepsilon_2^2)$. In Eq.~\eqref{eq:appendix_approximate_eta}, stochastic terms drop out as they are of higher oder $\propto O(\varepsilon_1,\varepsilon_2)$ than the desired accuracy.
Moreover, under the adiabatic assumption, $\mathcal{L}_0$ can be neglected as it is far smaller than the atomic internal dynamics characterized by $\Omega(\hat{z})$, $\Delta_t$ and $\Gamma_{e(t)}$. By keeping terms up to first order in $\Gamma_{e(t)}$ under the condition under the condition $\Gamma_t,\Gamma_e \ll\Omega(z),|\Delta_t|,|\Omega(z)/2\pm\Delta_t|$, the above equations can be solved adiabatically to give
\begin{align}
\hat{\eta}_{\pm}=&\frac{2}{\hbar}\Omega^{-2}(\hat{z})\left[\mp \Omega(\hat{z})-i\frac{\Gamma_e}{2}\right]\hat{H}_1\rho_c,\nonumber\\
\hat{\eta}_{t}=&\frac{1}{\Delta^2_t}\left(\Delta_t-i\frac{\Gamma_t}{2}\right)\tilde{g}(\hat{z})\hat{c}\rho_c.
\end{align}

Similarly, for $\hat{\zeta}_i$ we need to know their evolution accurate to second order in $\varepsilon_{1,2}$ for the deterministic terms and to linear order in $\varepsilon_{1,2}$ for the stochastic term, $d\hat{\zeta}_i=O(\varepsilon_1^2,\varepsilon_2^2)dt+O(\varepsilon_1,\varepsilon_2)dW(t)$. They are given by
\begin{align}
d\hat{\zeta}_t=&-\Gamma_t\hat{\zeta}_t+\mathcal{L}_0\hat{\zeta}_tdt-i[\tilde{g}(\hat{z})\hat{c}\hat{\eta}^\dag_t-\hat{\eta}_t\tilde{g}(\hat{z})\hat{c}^\dag]dt,\nonumber\\
d\hat{\zeta}_\pm=&-\frac{\Gamma_e}{2}\hat{\zeta}_\pm dt +\mathcal{L}_0\hat{\zeta}_\pm dt-\frac{i}{\hbar}(\hat{H}_1\hat{\eta}^\dag_\pm-\hat{\eta}_\pm\hat{H}_1)dt\nonumber\\
&+\frac{1}{2}\Big[\Gamma_t{\cal K}_2^{rt}\hat{\zeta}_t+\frac{\Gamma_e}{2}(P_{ge}\mathcal{K}_1^{ge}+P_{re}\mathcal{K}_2^{re})(\hat{\zeta}_++\hat{\zeta}_-)\Big]dt.
\end{align}
The first equation can be solved adiabatically to give
\begin{align}
\hat{\zeta}_t=&\frac{1}{\Delta_t^2}\tilde{g}(\hat{z})\hat{c}\rho_c\tilde{g}(\hat{z})\hat{c}^\dag.
\end{align}
In solving the second equation, we make an expansion in the branching ratios $P_{ge},P_{re}\ll 1$, and retain only first order terms in $P_{ge},P_{re}$. This will provide an effective evolution of  $\rho_c$ accurate to first order in $P_{ge},P_{re}$ [see the last line of Eq.~\eqref{darkstate_SME_1}]. We thus get
\begin{align}
\hat{\zeta}_{\pm}&=\frac{\Gamma_t}{\Gamma_e}(1+P_{ge}\mathcal{K}_1^{ge}+P_{re}\mathcal{K}_2^{re})\mathcal{K}_2^{rt}\hat{\zeta}_t\nonumber\\
&\pm\frac{4i}{\hbar^2\Gamma_e}\left[\hat{H}_1\rho_c\hat{H}_1\Omega^{-1}(\hat{z})-{\rm H.c.}\right]\nonumber\\
&+\frac{2}{\hbar^2}(1+P_{ge}\mathcal{K}_1^{ge}+P_{re}\mathcal{K}_2^{re})\left[\hat{H}_1\rho_c\hat{H}_1\Omega^{-2}(\hat{z})+{\rm H.c.}\right].
\end{align}

Finally, we note that ${\rm Tr}(\hat{c}\rho_c)={\rm Tr}(\hat{c}\mu_c)+O(\varepsilon_1^2,\varepsilon_2^2)$. Thus, with the accuracy prescribed by Eq.~\eqref{perturbativeExpression}, it is justified to express the quadrature in the nonlinear operator $\mathcal{H}[\hat{c}]$ in Eq.~\eqref{darkstate_SME_1} in terms of the reduced density matrix $\rho_c$, i.e. as $\langle\hat{c}\rangle_c={\rm Tr}(\hat{c}\rho_c)$.

Plugging these solutions to Eq.~\eqref{darkstate_SME_1}, we find
\begin{align}
\label{darkstate_SME_simplified}
d\rho_{c} =& \mathcal{L}_0 \rho_{c}dt  -\frac{i}{\hbar}[\hat{H}_{\rm coup},\rho_c]dt\nonumber\\& +  \sqrt{\kappa}\mathcal{H}[\hat{c}e^{-i\phi}]\rho_c dW(t)+\left(\frac{\Gamma_t}{\Delta_t^2}\mathcal{L}'+\mathcal{L}''\right)\rho_c dt,
\end{align}
Here $\mathcal{L}_0$  is defined in Eq.~\eqref{eq:L0_app}, $\mathcal{H}[\hat{c}]\rho_c=(\hat{c}-\langle \hat{c}\rangle_c)\rho_c + {\rm H.c.}$, with $\langle\hat{c}\rangle_c={\rm Tr}(\hat{c}\rho_c)$. While 
\begin{equation}
\label{eq:Hcoup_app}
\hat{H}_{\rm coup}=\frac{\hbar g^2(\hat{z})}{\Delta_t}\left[\cos\alpha(\hat{z})\right]^2\hat{c}^\dag\hat{c}
\end{equation}
is the desired local atom-cavity couping.
The Liouvillians $\mathcal{L}'$ and $\mathcal{L}''$ describes the effects of atomic spontaneous emission,
\begin{align}
\label{eq:LpLpp_app}
\mathcal{L}'\rho_c=&(\mathcal{G}_1^{rt}[\tilde{g}(\hat{z})\hat{c}]+\mathcal{G}_2^{rt}[\tilde{g}(\hat{z})\hat{c}])\rho_c\nonumber\\
&+(P_{ge}\mathcal{K}_2^{ge}+P_{re}\mathcal{K}_1^{re}-1)\mathcal{K}_2^{rt}\left(\tilde{g}(\hat{z})\hat{c}\rho_c\tilde{g}(\hat{z})\hat{c}^\dag\right),\nonumber\\
\mathcal{L}''\rho_c=&\frac{2\Gamma_e}{\hbar^2}(P_{ge}\mathcal{K}_2^{ge}+P_{re}\mathcal{K}_1^{re})[\hat{H}_1\rho_c\hat{H}_1\Omega^{-2}(\hat{z})+{\rm H.c.}]\nonumber\\
&-\frac{2}{\hbar^2}{\Gamma_e}\{\hat{H}_1\Omega^{-2}(\hat{z})\hat{H}_1,\rho_c\}.
\end{align}
We note $\mathcal{L}''$ can be suppressed indefinitely small independently of $\hat{H}_{\rm coup}$ by increasing the amplitude of $\Omega_g$ and $\Omega_r(z)$ while keeping their ratio fixed. In contrast, $\Gamma_t/\Delta_t^2\mathcal{L}'$ cannot be suppressed independent of $\hat{H}_{\rm coup}$ and constitutes an essential imperfection to the operation of the microscope. We thus retain $\Gamma_t/\Delta_t^2\mathcal{L}'$ and neglect $\mathcal{L}''$ in the main text. 

Using Eqs.~\eqref{eq:supG_appendix} and~\eqref{eq:supK_appendix}, we can write down the detailed form of $\mathcal{L}'$,
\begin{align}
\label{SPsuperOP1}
&\mathcal{L}'\rho_c  =\int duN_{rt}(u)\mathcal{D}\left[\hat{c}\,e^{-ik_{0}u\hat{z}}\hat{v}_{z_{0}}\right]\rho_c\nonumber\\
 & +P_{ge}\!\int \!dudu'N_{rt}(u)N_{ge}(u')\mathcal{D}\left[\hat{c}\,e^{-ik_{0}(u+u')\hat{z}}\hat{w}_{z_{0}}\sin\hat{\alpha}_{z_0}\right]\!\rho_c\nonumber\\
 &+P_{re}\!\int \!dudu'N_{rt}(u)N_{re}(u')\mathcal{D}\left[\hat{c}\,e^{-ik_{0}(u+u')\hat{z}}\hat{v}_{z_{0}}\sin\hat{\alpha}_{z_0}\right]\rho_c\nonumber\\
 &-\frac{1}{2}\Big\{ \hat{c}^{\dagger}\hat{c}\left[\hat{w}^{2}_{z_{0}}(1-P_{ge}\sin^2\hat{\alpha}_{z_0})-P_{re}\hat{v}^2_{z_{0}}\sin^{2}\hat{\alpha}_{z_0}\right],\,\rho_c\Big\}
\end{align}
Here we have used the shorthand notation $\hat{\alpha}_{z_0}\equiv\alpha(\hat{z})$, and have defined
\begin{align}
\hat{v}_{z_{0}}& =g(\hat{z})\cos^{2}\hat{\alpha}_{z_0},\nonumber\\
\hat{w}_{z_{0}} & =g(\hat{z})\cos\hat{\alpha}_{z_0}\sin\hat{\alpha}_{z_0}=\hat{v}_{z_{0}}\tan\hat{\alpha}_{z_0}.
\end{align}

\begin{figure}[t!]
\centering{}\includegraphics[width=0.5\textwidth]{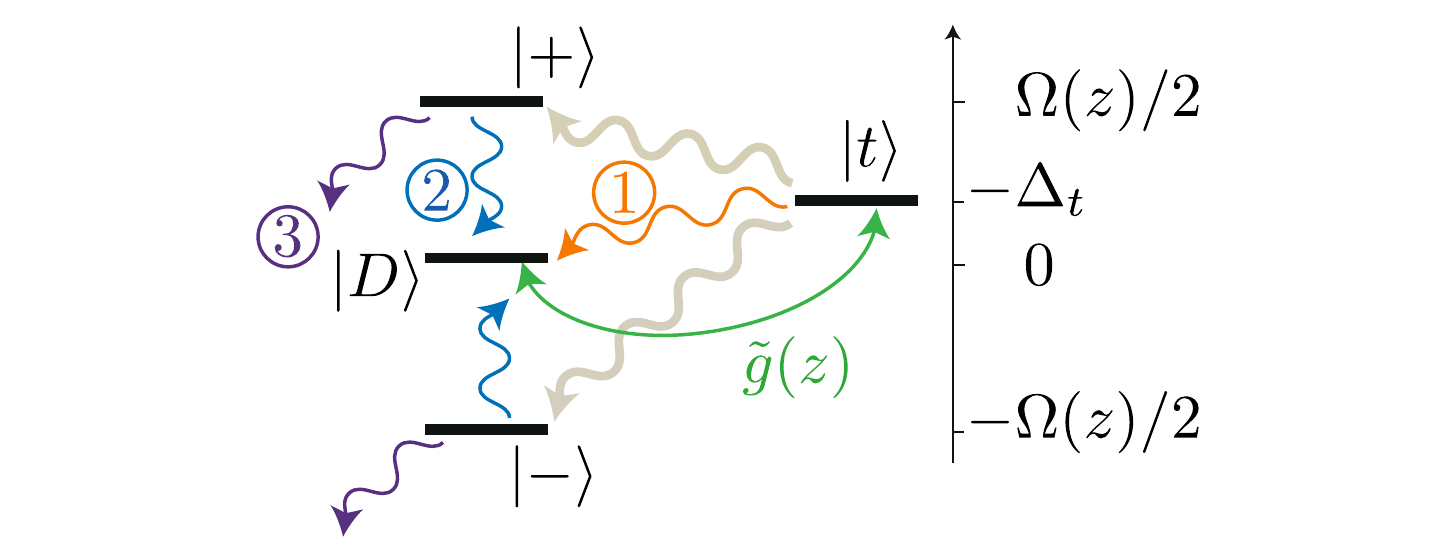} \caption{Atomic dressed states and the spontaneous emission channels relevant to Eq.~\eqref{SPsuperOP1}. The $\ket{D}\to\ket{t}$ transition interacts with the cavity mode with a strength $\tilde{g}(z)=-g(z)\cos\alpha(z)$, which leads to the local coupling Eq.~\eqref{eq:Hcoup_app} in second order perturbation theory. The wavy lines describe different spontaneous emission channels, see text for a detailed explanation.}
\label{fig:figureA1} 
\end{figure}

The terms in Eq.~\eqref{SPsuperOP1} are shown schematically in Fig.~\ref{fig:figureA1}. Here, the first line describes the process where the atom virtually absorbs a cavity photon and makes a transition from $\ket{D}$ to $\ket{t}$, then decays back to $\ket{D}$ while spontaneously emitting a photon into free space, shown as channel 1 in Fig.~\ref{fig:figureA1}. The second and third lines describe another possible route: after making a virtual transition from $\ket{D}$ to $\ket{t}$ by absorption of a cavity photon, the atom first spontaneously decays to the bright states $\ket{\pm}$, then undergoes a second spontaneous emission to go back to $\ket{D}$, shown as channel 2 in Fig.~\ref{fig:figureA1}. The fourth line is trace negative, and describes the decay of the atomic population due to the spontaneous emission from $\ket{\pm}$ to states outside the four-level system, shown as channel 3 in Fig.~\ref{fig:figureA1}.

Finally, the subwavelength condition $\sigma/\lambda_0\ll 1$ [cf. Sec.~\ref{sec:focusingFunction} of the main text] allows us to simplify Eq.~\eqref{SPsuperOP1}. Expanding Eq.~\eqref{SPsuperOP1} in power series of $\sigma/\lambda_0$, to the lowest order we can replace ${\rm exp}(-ik_0 u \hat{z})$ by ${\rm exp}(-ik_0 u z_0)$. Cosequently, the momentum diffusion effect in the first three lines of Eq.~\eqref{SPsuperOP1} can be neglected, and the mechanical effects on the atom are captured by the spatially localized operators $\hat{v}_{z_0},\hat{w}_{z_0}$ and $\hat{\alpha}_{z_0}$. We will adopt this approximated expression for $\mathcal{L}'$ when deriving the spontaneous-emission induced decoherence to the atomic external motion, Eq.~\eqref{eq:Lsp_appendix}.

\section{Calculation of the signal-to-noise ratio}
\label{sec:SNRcascade}
In this appendix we detail the technique we adopt for calculating the SNR of the filtered homodyne current [cf., Eqs.~\eqref{eq:filteredIh} and~\eqref{eq:SNRdef}]. While the SNR can in principle be extracted, according to its definition Eqs.~\eqref{eq:filteredIh} and~\eqref{eq:SNRdef}, by statistical averaging over all measurement trajectories, this is practically inefficient. Instead, specific for the filter function Eq.~\eqref{simpleFilter}, we can simplify the calculation by introducing an auxiliary cavity to the microscope setup as a physical filter. The SNR of the filtered homodyne current can thus be expressed in terms of the lower order moments of the auxiliary cavity mode, which can be calculated straightforwardly by solving the corresponding cascade master equation.

Given the microscope setup [cf. Fig.~\ref{fig:figure1} of the main text], we consider feeding its output field to an auxiliary cavity mode. We denote the destruction (creation) operator of this auxiliary mode as $\hat{c}_{\rm a}(\hat{c}_{\rm a}^\dag)$, and fix its frequency to be the same as the frequency of the microscope cavity. Moreover, we choose its linewidth as $\kappa_{\rm a}=2/\tau$, where $\tau$ is the filter integration time [cf. Eq.~\eqref{simpleFilter} of the main text]. From such a construction, the auxiliary cavity mode serves as a physical filter of the output field of the microscope. This is revealed most directly in the Heisenberg picture, in which the evolution of the auxiliary mode reads
\begin{equation}
\dot{\hat{c}}_{\rm a}(t) =  - \frac{1}{\tau}\hat{c}_{\rm a}(t)-\sqrt{\frac{2}{\tau}}\hat{f}_{\rm out}(t),
\label{eq:auxiliaryDynamics}
\end{equation}
where $\hat{f}_{\rm out}(t)$ is the output field of the microscope. Eq.~\eqref{eq:auxiliaryDynamics} can be integrated straightforwardly
\begin{equation}
\hat{c}_{\rm a}(t)=-\sqrt{\frac{2}{\tau}}\int_0^{\infty} d t' e^{-t'/\tau}\hat{f}_{\rm out}(t-t'). 
\label{eq:auxiliaryDynamicsIntegrated}
\end{equation}
Comparing Eq.~\eqref{eq:auxiliaryDynamicsIntegrated} to Eqs.~\eqref{eq:filteredIh} and \eqref{simpleFilter} of the main text, we find that 
\begin{align}
\langle \mathcal{I}_{\tau}(t) \rangle_{\rm st} &= - \langle \hat{X}_{\rm a}^{\phi}(t)\rangle/\sqrt{2\tau},\nonumber\\
\langle \delta\mathcal{I}^2_{\tau}(t) \rangle_{\rm st} &=  \langle [\delta\hat{X}_{\rm a}^{\phi}(t)]^2\rangle / 2\tau.
\label{eq:auxiliary_statistics}
\end{align}
Here, we have defined the quadrature operator of the auxiliary cavity mode in the Heisenberg picture, $\hat{X}_{\rm a}^{\phi}(t)=e^{-i\phi}\hat{c}_{\rm a}(t) + e^{i\phi}\hat{c}_{\rm a}^{\dag}(t) $, and its fluctuation $\delta\hat{X}_{\rm a}^{\phi}(t)=\hat{X}_{\rm a}^{\phi}(t)-\langle\hat{X}_{\rm a}^{\phi}(t)\rangle$, where $\langle\dots\rangle\equiv{\rm Tr}[\varrho_0\dots]$ is an expectation value with respect to the initial density matrix of the microscope plus the auxiliary-cavity.
Thus, the statistics (and thus the SNR) of the filtered homodyne current is directly imbedded in the lower-order moments of the auxiliary cavity mode.

In the above we adopt the Heisenberg picture to arrive at Eq.~\eqref{eq:auxiliary_statistics}. Nevertheless, to calculate Eq.~\eqref{eq:auxiliary_statistics} it is more convenient to adopt the Schr\"odinger picture. In this picture, the RHS of Eq.~\eqref{eq:auxiliary_statistics} can be expressed as $\langle\hat{X}_{\rm a}^{\phi}(t)\rangle={\rm Tr}[\hat{X}_{\rm a}^{\phi}\varrho(t)]$ and $\langle[\delta\hat{X}_{\rm a}^{\phi}(t)]^2\rangle={\rm Tr}[(\hat{X}_{\rm a}^{\phi})^2\varrho(t)]-\{{\rm Tr}[\hat{X}_{\rm a}^{\phi}\varrho(t)]\}^2$, where $\hat{X}_{\rm a}^{\phi}=e^{-i\phi}\hat{c}_{\rm a} + e^{i\phi}\hat{c}_{\rm a}^{\dag}$, and $\varrho(t)$ is the density matrix of the microscope plus the auxiliary-cavity. It evolves according to the cascade master equation
\begin{align}
\label{cascadeME}
\dot{\varrho}  = &-\frac{i}{\hbar}[\hat{H}_{A,E}+\hat{H}_C+\hat{H}_{\rm coup},\varrho]dt+\kappa \mathcal{D}[\hat{c}]\varrho dt+\frac{\Gamma_t}{\Delta_t^2}\mathcal{L}'\varrho dt\nonumber\\
&+\frac{2}{\tau}\mathcal{D}[\hat{c}_{\rm a}]\varrho-\sqrt{\frac{2\kappa}{\tau}}\Big( [\hat{c}^{\dag}_{\rm a}, \hat{c}\varrho] + [\varrho\hat{c}^\dag, \hat{c}_{\rm a}]\Big).
\end{align}
The first line of Eq.~\eqref{cascadeME} corresponds to the unconditional dynamics of the microscope setup [cf. Eq.~\eqref{eq:SME_firstElimination}], with small terms $\propto V_{\rm na}(\hat{z})$ and $\propto \mathcal{L}''$ neglected. The second line corresponds to the dynamics of the auxiliary cavity mode, which acts as a physical of filter of the homodyne current.

By numerically propagating Eq.~\eqref{cascadeME}, we extract the statistics of the filtered homodyne current and thus the SNR.

\section{Perturbative elimination of the cavity mode}
\label{adiabaticEliminationCavity}
In this appendix we derive the stochastic master equations~\eqref{eq:SME_elimination_badcavity} and~\eqref{eq:SME_good_cavity} together with the corresponding homodyne currents~\eqref{eq:homodyne_elimination_badcavity} and \eqref{eq:homodyne_good_cavity_maintext}, by eliminating the cavity mode perturbatively starting from Eqs.~\eqref{eq:SME_firstElimination} and~\eqref{eq:Ih_firstElimination} of the joint atom-cavity system. This will allow us to relate the homodyne current to effective observables of the atom in the bad/good cavity limit, thus to define the two operation modes of the microscope.

We start by shifting away the stationary amplitude of the cavity field. Without coupling to the atom, the driven cavity mode populates a coherent state with amplitude $\alpha_0=-2\mathcal{E}/\sqrt{\kappa}$ (we assumed $\alpha_0$ being real and $\alpha_0\gg 1$ hereafter). We can shift it away via the transformation $\rho_c\to\hat{U}(\alpha_0)\rho_c\hat{U}(-\alpha_0)$, $\hat{H}\to\hat{U}(\alpha_0)\hat{H}\hat{U}(-\alpha_0)$ etc., with the unitary operator $\hat{U}(\alpha_0)={\rm exp}(\alpha_0 \hat{c}-\alpha_0 \hat{c}^\dag)$. As a result, Eq.~\eqref{eq:SME_firstElimination} of the main text is transformed into
\begin{eqnarray}
\label{eq:SME_firstElimination_appendix}
d\rho_c=&&-\frac{i}{\hbar}[\hat{H}_{A,E}+\hat{H}_{\rm coup}',\rho_c]dt+\kappa \mathcal{D}[\hat{c}]\rho_{c}dt\nonumber\\
&&+\sqrt{\!\kappa}\mathcal{H}[\hat{c}e^{\!-\!i\phi}]\rho_{c} dW\!(t) + \frac{\gamma}{4\mathcal{C}}\mathcal{L}_{\rm sp}\rho_c dt.
\end{eqnarray}
Here $\hat{H}_{\rm coup}'=\mathcal{A}f_{z_0}(\hat{z})(\hat{c}^\dag\hat{c}+\alpha_0\hat{c}^\dag+\alpha_0\hat{c})\simeq\alpha_0\mathcal{A}f_{z_0}(\hat{z})(\hat{c}^\dag+\hat{c})$, and $\mathcal{L}_{\rm sp}$ is given by the replacement $\hat{c}\to\alpha_0$ in the expression of $\mathcal{L}'$ [see Eq.~\eqref{SPsuperOP1}],
\begin{align}
&\mathcal{L}_{{\rm sp}}\rho =\vphantom{\int}\mathcal{D}\left[f_{z_{0}}(\hat{z})\right]\rho+P_{r}\mathcal{D}\left[f_{z_{0}}(\hat{z})\sin\hat{\alpha}_{z_0}\right]\rho\nonumber\\
&+P_{g}\mathcal{D}\left[f_{z_{0}}(\hat{z})\tan\hat{\alpha}_{z_0}\sin\hat{\alpha}_{z_0}\right]\rho\nonumber\\
 &-\frac{1}{2}\left\{ f_{z_{0}}^{2}(\hat{z})[\tan^{2}\hat{\alpha}_{z_0}(1-P_{g}\sin^{2}\hat{\alpha}_{z_0})-P_{r}\sin^{2}\hat{\alpha}_{z_0}],\rho\right\}
 \label{eq:Lsp_appendix}
\end{align}
In Eq.~\eqref{eq:SME_firstElimination_appendix} we have dropped an optical lattice potential $V_{\rm OL}(\hat{z})=\mathcal{A}f_{z_0}(\hat{z})\alpha_0^2$ due to the stationary cavity field, as it can be compensated straightforwardly by detuning the Raman resonance with a small offset $\Delta_r=g^2(z_0)\alpha_0^2/\Delta_t$. We have also neglected the small decoherence terms in $\mathcal{L}_{\rm sp}$ involving the fluctuations of the cavity field. Eq.~\eqref{eq:Ih_firstElimination} is transformed to
\begin{equation}
\label{eq:homodyneCurrent_appendix}
I(t)dt = \sqrt{\kappa}\langle \hat{c} e^{-i\phi} +\hat{c} ^\dagger e^{i\phi}\rangle_c\,dt +dW(t),
\end{equation}
where $\hat{c}(\hat{c}^\dag)$ now corresponds to the fluctuation of the cavity field, and we have dropped a constant term contributed by the stationary cavity field.

Next we move into the interaction picture with respect to $\hat{H}_{A,E}$, the Hamiltonian for the atomic external motion. As a result, Eq.~\eqref{eq:SME_firstElimination_appendix} is transformed into
\begin{eqnarray}
\label{eq:SME_firstElimination_rotatingframe_appendix}
d\rho_c=&&-\frac{i}{\hbar}\mathcal{A}\alpha_0[\hat{f}_{z_0}(t)(\hat{c}+\hat{c}^\dag),\rho_c]dt+\kappa \mathcal{D}[\hat{c}]\rho_{c}dt\nonumber\\
&&+\sqrt{\!\kappa}\mathcal{H}[\hat{c}e^{\!-\!i\phi}]\rho_{c} dW\!(t) + \frac{\gamma}{4\mathcal{C}}\mathcal{L}_{\rm sp}(t)\rho_c dt.
\end{eqnarray}
Here, we have defined in the interaction picture $\hat{f}_{z_0}(t)\equiv{\rm exp}(i\hat{H}_{A,E}t/\hbar)f_{z_0}(\hat{z}){\rm exp}(-i\hat{H}_{A,E}t/\hbar)$. We have also defined a time-dependent decoherence term, $\mathcal{L}_{\rm sp}(t)$, of which the expression can be yielded  by replacing $v_{z_0}(\hat{z})$ and $w_{z_0}(\hat{z})$ in Eq.~\eqref{eq:Lsp_maintext} by the corresponding operators in the interaction picture. The expression for the homodyne current Eq.~\eqref{eq:homodyneCurrent_appendix} remains the same under this transformation.

Now we eliminate the cavity mode under the condition of weak atom-cavity coupling $\mathcal{A}\alpha_0\ll \hbar\kappa$, by perturbation theory in the small parameter $\varepsilon=\mathcal{A}\alpha_0/(\hbar\kappa)$. We define a conditional density matrix for the atomic external motion, $\tilde{\rho}_c\equiv{\rm Tr}_C(\rho_c)$, where ${\rm Tr}_{C}$ stands for tracing over the states of the cavity mode.
Our aim is to derive an equation for the evolution of $\tilde{\rho}_c$ up to second order in the perturbation $\varepsilon$ and up to a linear stochastic term
\begin{equation}
\label{perturbationTheory2}
d\tilde{\rho}_c=O(\varepsilon^2)dt+O(\varepsilon)dW(t).
\end{equation}
To this end, we trace out the cavity DOF in Eq.~\eqref{eq:SME_firstElimination_rotatingframe_appendix}, yielding
\begin{align}
d\tilde\rho_{c} & =-\frac{i}{\hbar}\mathcal{A}\alpha_0\!\left[\hat f_{z_0}(t),\hat \eta\!+\!\hat \eta^{\dagger}\right]dt+\sqrt{\kappa}\left(\hat \varsigma e^{-i\phi}\!+\!{\rm h.c.}\right)dW\!(t).
\label{eq:rho_S}
\end{align}
Here $\hat{\eta}={\rm Tr}_C(\hat{c}\rho_c)$ and $\hat{\varsigma}=\hat{\eta}-{\rm Tr}_S(\hat{\eta})\tilde{\rho}_c$, where ${\rm Tr}_S$ stands for tracing over the states of atomic external motion. The homodyne current can be related to $\hat{\eta}$ as
\begin{equation}
\label{eq:homodyne_elimination_app}
dq(t) \equiv I(t)dt = \sqrt{\kappa}\Big[{\rm Tr}_S(\hat{\eta})e^{-i\phi}+{\rm c.c.}\Big]dt + dW(t).
\end{equation}
To derive an equation for $\tilde{\rho}_c$ with the accuracy prescribed by Eq.~\eqref{perturbationTheory2}, we need to solve for $\hat \eta$ and $\hat \varsigma$ accurate to first order in the perturbation $\varepsilon$ and up to a zeroth order stochastic term. With this accuracy, their evolution is given by
\begin{align}
d\hat{\eta}&=-\frac{i}{\hbar}\mathcal{A}\alpha_0\hat{f}(t)\tilde{\rho}_cdt-\frac{\kappa}{2}\hat{\eta}dt,\nonumber \\
d\hat{\varsigma}&=-\frac{i}{\hbar}\mathcal{A}\alpha_0[\hat{f}(t)-\langle\hat{f}(t)\rangle_c]\tilde{\rho}_c dt -\frac{\kappa}{2}\hat{\varsigma}dt,
\label{2ndEliminationFirstOderEqns}
\end{align}
where we have neglect terms involving $\mathcal{L}_{\rm sp}(t)$ in view of the smallness of the energy scale of $\mathcal{L}_{\rm sp}(t)$ compared to the cavity damping $\kappa$.
By solving Eq.~\eqref{2ndEliminationFirstOderEqns} adiabatically and plugging these solutions back to Eq.~\eqref{eq:rho_S}, we can derive an effective equation of motion for $\tilde{\rho}_c$. In the following we analyze the example of measuring a harmonically trapped atom, and derive the effective equation of motion corresponding to the bad/good cavity limit separately.
\subsubsection{Bad cavity limit}
The bad cavity limit is defined by $\kappa\gg \omega$, i.e., the cavity dynamics is much faster than the atomic motion (quantified by its oscillation frequency $\omega$), such that the former instantaneously follows the latter. Eq.~\eqref{2ndEliminationFirstOderEqns} can thus be solved adiabatically
\begin{align}
\hat{\eta}&=-2i \alpha_0\frac{\mathcal{A}}{\hbar\kappa}\hat{f}(t)\tilde{\rho}_c,\nonumber\\
\hat{\varsigma}&=-2i\alpha_0\frac{\mathcal{A}}{\hbar\kappa}[\hat{f}(t)-\langle\hat{f}(t)\rangle_c]\tilde{\rho}_c.
\end{align}
Plugging these solutions back to Eqs.~\eqref{eq:rho_S} and~\eqref{eq:homodyne_elimination_app}, restoring the Schr\"odinger picture, and choose the homodyne angle $\phi=-\pi/2$ to maximize the homodyne current, we arrive at \eqref{eq:SME_elimination_badcavity} and \eqref{eq:homodyne_elimination_badcavity} in the main text.
\subsubsection{Good cavity limit}
The good cavity limit is defined by $\kappa\ll \omega$, i.e., the cavity dynamics is much slower than the atomic motion. To solve Eq.~\eqref{2ndEliminationFirstOderEqns} in this limit, we expand the time-evolving focusing function in terms of its sidebands, $\hat{f}_{z_0}(t)=\sum_{\ell}\hat{f}_{z_0}^{(\ell)}e^{-i\ell\omega t}$, with the $\ell$-th \emph{sideband component} $\hat f_{z_{0}}^{(\ell)}=\sum_{n}f_{n,n+\ell}|n\rangle\langle n+\ell|$, and $f_{mn}=\langle m|f_{z_{0}}(\hat{z})|n\rangle$. Assuming that $\tilde{\rho}_c$ depends on time slowly, Eq.~\eqref{2ndEliminationFirstOderEqns} can be integrated as
\begin{align}
\hat{\eta}&=-\frac{i}{\hbar}\mathcal{A}\alpha_0\sum_{\ell}\frac{\hat{f}^{(\ell)}_{z_0}e^{-i\ell\omega t}}{\kappa/2-i\ell\omega}\tilde{\rho}_c,\nonumber\\
\hat{\varsigma}&=-\frac{i}{\hbar}\mathcal{A}\alpha_0\sum_{\ell}\frac{(\hat{f}^{(\ell)}-\langle\hat{f}^{(\ell)}\rangle_c)}{\kappa/2-i\ell\omega}e^{-i\ell\omega t}\tilde{\rho}_c.
\end{align}
Substituting these expression into Eq.~\eqref{eq:rho_S}, restoring the Schr\"odinger picture and keeping only the non-rotating terms in the rotating-wave approximation, we obtain
\begin{align}
d\tilde{\rho}_{c} =&-\frac{i}{\hbar}[\hat{H}_{{\rm eff}},\tilde{\rho}_{c}]\,dt+\sum_{\ell}\frac{\gamma}{1+(2\omega\ell/\kappa)^{2}}{\cal D}[\hat f_{z_{0}}^{(\ell)}]\tilde{\rho}_{c}\,dt\nonumber \\
 & +\sqrt{\gamma}\sum_{\ell}{\cal H}\Big[\frac{\hat f_{z_{0}}^{(\ell)}}{1-2i\ell\omega/\kappa}\Big]\tilde{\rho}_{c}\,dW(t)+\frac{\gamma}{4\mathcal{C}}\mathcal{L}_{\rm sp}\tilde{\rho}_cdt,\label{eq:SME_good_cavity_elimination-1}
\end{align}
where 
\begin{equation}
\label{eq:app_Heff}
\hat{H}_{\rm eff}=\hat{H}_{S}+\sum_{\ell}\frac{\alpha_0^2\mathcal{A}^2\ell\omega/\hbar}{(\kappa/2)^2+\ell^2\omega^2}\Big[\hat{f}^{(\ell)}\hat{f}^{(\ell)\dag}-\hat{f}^{(\ell)\dag}\hat{f}^{(\ell)}\Big],
\end{equation}
and we have chosen the homodyne angle $\phi=-\pi/2$ to enhance the signal $\propto\langle \hat{f}_{z_0}^{(0)}\rangle_c$. By filtering out higher sideband components corresponding to $\ell\neq 0$ in the homodyne current with a classical filter, the homodyne current can be expressed as
\begin{equation}
\label{eq:homodyne_good_cavity}
dq(t)\equiv I(t)dt=2\sqrt{\gamma}\,\langle \hat f_{z_{0}}^{(0)}\rangle_{c}\,dt+dW(t),
\end{equation}
which recovers Eq.~\eqref{eq:homodyne_good_cavity_maintext} of the mian text. Since these higher sideband components are not resolved, we can drop the stochastic terms with $\ell\neq 0$ in Eq.~\eqref{eq:SME_good_cavity_elimination-1}, corresponding to averaging over these unobserved measurement channels. Finally, in the good cavity limit $\kappa\ll\omega$, the second term in the Hamiltonian $\hat{H}_{\rm eff}$ [cf. Eq.~\eqref{eq:app_Heff}] is much smaller than $\hat{H}_{S}$ and can be neglected. We thus arrive at Eq.~\eqref{eq:SME_good_cavity} of the main text.
\end{appendix}
\end{document}